# The Effects of Information and Communication Technology Use on Human Energy and Fatigue: A Review

Jana Korunovska[1,2] and Sarah Spiekermann[1]

*Information and communication technologies (ICTs) are generally assumed to save time and energy, yet user fatigue due to ICT use is on the rise. The question about the effects of ICT use on human energy and fatigue is marred by inconsistencies in terminology, definitions, and measures. The aim of this paper is therefore twofold. First, we provide a consolidation and classification of subjective energy and fatigue concepts from four leading research areas. Second, we review the empirical literature on the relationship between ICT use and seven different subjective energy and fatigue concepts from the four areas. We show that ICT use can both energize and fatigue users, sometimes even at the same time, a phenomenon that we term Digital Fatigue Paradox. Overall, there is more evidence for the fatiguing effect, which also appear to be stronger, even though ICT users might actually believe the opposite to be true. By consolidating the mechanisms through which ICT use energizes and fatigues users in a conceptual model, we provide initial explanation for the paradox and derive implications for organizational policy, ICT design, and regulation that strive to improve the user experience with ICTs and prevent ill-being, i.e., foster well-being.*

## INTRODUCTION

New information and communication technologies (ICTs) were invented to make humanity more efficient and productive, save us time and energy, and improve our quality of life and well-being. Yet lack of time and energy is the number one complaint of modern society, and digital technology is often seen as a cause. The number of employees who suffer from fatigue and emotional exhaustion is rising and often exceeds 25% of the general working population, depending on country and assessment method (Aumayr-Pintar et al. 2018; Bültmann et al. 2002; Shanafelt et al. 2019). The cost of burned-out employees is estimated to be up to US$190 billion per year in healthcare spending in the US alone (Han et al. 2019).

Digitalization of the workplace and distraction through smartphones and Social Networking Sites (SNS) have been cited as reasons for employees' exhaustion and high productivity loss (Ayyagari et al. 2011; Bialowolski et al. 2020). Consequently, terms such as "techno-stress," "tech-invasion," and "digital fatigue" have entered the vocabulary (Gaudioso et al. 2017; Ragu-Nathan et al. 2008). In addition, "Internet fatigue" and exhaustion through SNS addiction extend the problem beyond the working population (Lin et al. 2013). Especially affected by this problem are younger

---

[1] Vienna University of Economics and Business, Austria; [2] Corresponding author: jana.korunovska [at] wu.ac.at

Internet and SNS users who are now frequently called "the burned-out generation" (Bener et al. 2018; Pendell 2018).

Despite these warning signs, investments in ICTs and digitalization continue to soar. Adoption and use of ICTs is steadily rising and has surpassed half of the world's population, with the numbers getting another significant push in the wake of the Covid-19 pandemic (Chaffey, 2020). So an important question arises: Are ICTs saving us energy or are they causing fatigue? Answering this question is difficult. Research on human energy and fatigue is marred by inconsistencies in terminology, definitions, and measures since researchers often use the same term to refer to different constructs or different terms to refer to the same one (Ackerman 2011; O'Connor 2004; Quinn et al. 2012).

For example, organizational scholars often focus on employees' vigor (Bakker et al. 2008)—but vigor as a construct is also found in the affective sciences, albeit with a different definition (McNair et al. 1989). Similarly, what motivation and social psychologists refer to as "vitality" (Deci and Ryan 2011) is different from vitality as referred to by medical psychologists (Ware Jr and Sherbourne 1992). Still, the constructs "vitality," "vigor-energy," and "energetic activation" are sometimes measured on the same adjective instrument with only slight variations in the scaling (O'Connor 2004).

On top of these inconsistencies, a multitude of other terms, often with indiscernible meaning or lacking operationalization, which nonetheless describe the subjective experiences of energy or fatigue, such as mental fatigue (Boksem and Tops 2008), mental energy (Lieberman 2007), emotional energy (Collins 1990), psychic energy (Csikszentmihalyi 1991), etc., are also often found in the literature. This inconsistency has spread to the Information Systems (IS) field, where IS researchers have borrowed and adopted constructs from many different disciplines and used them interchangeably. As a result, current knowledge on the ICT-use effects on energy and fatigue is scattered. Empirical insights are hardly comparable and frequently cause confusion.

In addition to this confusion and multiplicity of terms, research is "plagued" by the frequent misconception that feeling energetic is the opposite of feeling fatigued (McNair 1984, p.20). The timely "bivariate" paradigm holds that subjective energy, which is a positive affective state, and subjective fatigue, which is a negative affective state, are two separate experiences that can co-exist in humans (Mäkikangas et al. 2014). In other words, they are separate affective states that can be experienced simultaneously. Subjective energy and fatigue have different antecedents and behavioral consequences (Shirom 2011) as well as distinct neuronal correlates (Takeuchi et al. 2017). Despite this state-of-the-art knowledge about the human experience of energy and fatigue, the intuitive but problematic legacy of a "univariate paradigm" persists. The univariate paradigm sees energy and fatigue as two sides of the same coin, where the presence of one attests to the absence of the other. This univariate legacy has also permeated the IS discipline as most IS researchers focus either only on energy or only on fatigue. As a result, IS researchers frequently come to one-sided conclusions.

Against the background of these terminological and conceptual challenges, it becomes clear how complex it is to reliably derive generalizable conclusions about the effects of ICT use on human energy and fatigue. At this point in time, it is not possible to present a meta-analysis of the existing findings across constructs, considering the irreconcilable differences between their definitions and operationalization. But because the effects of ICT use on human energy and fatigue are so important for corporate IT investment, personal use decisions as well as political policy making, we still attempt in this paper to review what 42 empirical studies have found on the subject. For this purpose, we carefully observe the reported effects and mechanisms of ICT use on different subjective energy and fatigue constructs, thus controlling for both terminological ambivalence, and the pitfall of univariate measurement.

In order to facilitate conclusions and master the complexity of the widely diffused terminology we grouped the existing research into the four areas from which the terms originate. We then integrated the findings for each of these four areas with the help of seven construct categories, which denote either the experience of energy or that of fatigue:

(1) The mood states fatigue and vigor from the affective sciences and psychophysiology, which for the purposes of clarity we refer to as fatigue[Mood] and vigor[Mood].

(2) Work-related emotional exhaustion and vigor from organizational sciences and occupational psychology, in the following referred to as exhaustion[Work] and vigor[Work].

(3) Motivation-related subjective vitality from motivation and social psychology, in the following referred to as vitality[Mot].

(4) SNS-related exhaustion and fatigue from the fields of Information Systems (IS) and Human Computer Interaction (HCI), in the following referred to as exhaustion[SNS] and fatigue[SNS].

While integrating the widely diffused energy-related constructs under these seven summarizing category labels, we made sure that we only grouped those constructs together that have been measured in a similar way at the item level.

In the following section, we first discuss the details of our categorization, summarized in Table 1. The categorization is the first major contribution of this paper. It allows us to achieve the missing overarching perspective on the overall effects of ICT use on human energy and fatigue and we hope that it will be used in future research so that results can become comparable. We then report the findings of our systematic review based on this high-level category perspective. In what is the second contribution of the paper, we find that while ICT use can energize people, it is far more likely that it causes fatigue, often at the same time, a phenomenon we term the Digital Fatigue Paradox. By summarizing the patterns through which ICT use energizes and fatigues users, we arrive at our third major contribution, which is an integrated conceptual model that moves us towards a general theory on the relationship between ICT use and human energy and fatigue. Based on this model, we discuss the ICT design and policy implications that follow from our review and point towards the gaps in research that need to be filled for a more comprehensive understanding of the topic.

| Table 1. Summary and classification of subjective energy and subjective fatigue constructs | | | | | | | | |
|---|---|---|---|---|---|---|---|---|
| **Area** | **High-level construct** | **Original constructs** | **Main definition** | **Main measures** | **Duration** | **Range** | **Specifics** | **Referent population** |
| Affective Science Psychophysiology | Vigor$^{Mood}$ (N=8) | Vigor, energy, general activation, energetic activation, arousal, vitality, etc. | Feeling vigorous, energetic, active, full of pep, vital, lively. | Vigor-energy subscale of POMS | Short: momentary and up to a week | Narrow: only the feeling of energy | No specifics | General population |
| | | | | General activation subscale of AD ACL | Short: momentary feeling | | | |
| | Fatigue$^{Mood}$ (N=6) | Fatigue, mental fatigue deactivation, depletion, etc. | Feeling fatigued, worn-out, exhausted, weary, tired, depleted. | Fatigue subscale of POMS | Short: momentary and up to a week | Narrow: only the feeling of fatigue | No specifics | |
| | | | | Deactivation subscale of AD ACL | Short: momentary feeling | | | |
| Organizational and Occupational Science | Vigor$^{Work}$ (N=8) | Vigor, work vigor, engagement | Feeling high levels of energy and vigor *at work,* feeling resilient and persevering while working, looking forward to work | Vigor subscale of the UWES | Long lasting feeling | Very broad: next to the feeling of energy, it also includes mental resilience, will to invest energy | Felt at work, during work, and for work | Employees |
| | Exhaustion$^{Work}$ (N=16) | Emotional exhaustion, burnout, strain | Feeling fatigued, burned out and emotionally drained *by work*, feeling stressed from work | Emotional exhaustion subscale of the MBI | Long lasting, chronic feeling | A bit broader: next to the feeling of fatigue, it also includes feeling of work strain | Caused by work | |

| Area | High-level construct | Original constructs | Definition | Main measures | Duration | Range | Specifics | Referent population |
|---|---|---|---|---|---|---|---|---|
| Motivational psychology / Social psychology | Vitality[Mot] (N=4) | Subjective vitality | Experience of possessing life energy and aliveness that *comes from the self* | Subjective Vitality Scale (SVS) | Long lasting | A bit broader: next to the feeling of energy also feeling optimism, for the future | Achieved through continuous satisfaction of the basic psycho-logical needs, especially the need for autonomy | General population |
| Information Systems / Human Computer Interaction | Exhaustion[SNS] (N=8) | SNS exhaustion, Techno-exhaustion, SNS burnout, strain | Feeling exhausted, tired, burned out and drained *by SNS use* | SNS exhaustion scale (indirectly adopted from the MBI) | Longer lasting | A bit broader: next to the feeling of fatigue, it also includes feeling strain from SNS use | Caused by SNS use | SNS users |
|  | Fatigue[SNS] (N=4) | SNS Fatigue, Social network fatigue, Social media fatigue, mobile messenger fatigue | Feeling tired *from SNS use.* Feeling bored by, disinterested in, and indifferent to SNSs | Adopted mental fatigue scales. Various self-constructed scales | Short: during and immediately after SNS use | Very broad: next to the feeling of fatigue also feeling in-difference to SNS use, boredom caused by SNS use, and disinterest towards SNS use | Caused by SNS use |  |

Note. POMS = Profile of Mood States (POMS; McNair et al. 1971; Shacham 1983); AD ACL = Activation Deactivation Adjective Checklist (AD ACL; Thayer 1986); MBI = Maslach Burnout Inventory (Maslach et al. 1996; Maslach et al. 1986; Maslach et al. 2001)
UWES = Utrecht Work Engagement Scale (Schaufeli et al. 2006); SVS = Subjective vitality scale (Ryan and Frederick 1997); Social Networking Sites (SNS) exhaustion scale (Maier et al. 2015).

Energy is the capacity to work or the performance of work. In humans, molecular bonds derived from nutrition provide energy, whereas the neural arousal systems regulate it. During mental or physical work, potential energy available in molecular bonds (e.g., adenosine triphosphate (ATP), phosphocreatine (PCr), lactose, glycogen or blood glucose), is converted to biological work, such as neuronal activity or muscle contraction (Jones 2003; Magistretti and Allaman 2015; McArdle et al. 1991). These biological processes are not to be confused with the *experience* of having energy or being fatigued, yet many researchers do (Quinn et al. 2012, p.341).

The experience of having energy is indeed based on and closely related to the metabolic processes that transform conserved energy into our capacity to work, as well as to the activation of the different neural arousal systems such as the sleep-wake cycle (Jones 2003; McArdle et al. 1991). However, this relationship is far from perfect and is not yet very well established (Berrios 1990; Lieberman 2007; Marks 1977). Most of us experience fatigue even when we have a lot of energy as measured by physiological or neural correlates. We also experience situations that energize us even when we are physically exhausted. In other words, when people judge their energy or fatigue levels they do not assess their ATP, PCr or glucose levels. Instead, they judge their *feelings* about having energy or being fatigued, i.e., their *feelings* about investing energy in an activity or being fatigued as a result of it (Quinn et al. 2012). In this review, we focus solely on this *subjective experience* of having energy or being fatigued because this is what matters in the *user experience* (UX) with ICT use. In order to distinguish it from the biological energy potential and neural activation, and in line with tradition, we term the experience of having energy or being fatigued as subjective energy, i.e., subjective fatigue.

### The Subjective Energy and Fatigue Constructs Conceptualized as Mood

The most comprehensive research on subjective energy and subjective fatigue comes from the affective sciences. This is because the feelings, i.e., experiences of energy or fatigue are affective states (O'Connor 2006). Affective states themselves are commonly divided into *emotions* and *moods*. Emotions are usually specific, very short, intense states with a known cause. For example, feeling angry about something. Moods, on the other hand are more generalized, typically longer, fluctuating and less intense. They are usually unrelated to specific causes. For example, feeling blue for a while without apparent reason. Subjective energy can be experienced as an emotion towards something but it is most often regarded as an underlying core mood (O'Connor 2006; Quinn et al. 2012; Shirom 2011; Thayer 1990). Even though the number of the core dimensions of mood and their labeling have been disputed since the birth of psychology, the energy dimension reoccurs in most theories.

Subjective energy and fatigue are regarded as a fundamental underlying characteristic of all affective experiences (Kuppens et al. 2013). That said, the labels that are given to them differ from theory to theory. Prominent labels are energetic activation-deactivation or arousal (Thayer 1986), vigor-fatigue (McNair et al. 1971), vitality-fatigue (Ware Jr and Sherbourne 1992), etc. What allows us to identify these distinct labels as sharing the common subjective energy and fatigue core is that they use the same or highly similar measurement items. Typically, the items assess how a person is feeling or has been feeling with the help of adjectives that describe the experience of energy or fatigue. For example, "At the moment I feel energetic, vigorous, full of pep," etc. Or, "In the past week I have been feeling worn-out, tired, fatigued, exhausted," etc. One difference in these measures is a slight variation in the accessed duration of the mood, i.e., asking whether the mood is present right now or whether it has been present in the past week. The second difference is the choice and number of the synonymous adjectives.

Due to the overwhelming similarity of operationalization, we argue that the findings on subjective energy and subjective fatigue measured by various affect scales are similar enough to be grouped together regardless of the different labels that have been given to them across theories. As long as energy or fatigue are captured through highly similar adjective-scales, we group these affective states under the umbrella terms vigor[Mood] and fatigue[Mood].

What is not as obvious and consistent from the different measurements is how a particular affect theory conceptualizes the relationship between vigor[Mood] and fatigue[Mood]. Perhaps the most popular way to measure mood is a combination of arousal or activation with valence or pleasantness (Russell 1980), and often such measurement is conducted on a pictorial scale (Bradley and Lang 1994): This means that the energy mood, mirrored in the arousal scale is conceptualized as a univariate construct, i.e., as the opposite of fatigue or tiredness. Another popular way is to measure vigor[Mood] and fatigue[Mood] on two separate adjective subscales, which would imply a bivariate conceptualization (McNair et al. 1989; Thayer 1986). Though the authors of these prominent instruments believed that energy and fatigue are two end-poles of one dimension, they still used separate scales because this is what their empirical research suggested, i.e., their data did not fit on a single energy-fatigue dimension (McNair 1984). Most of these mood theories and instruments emerged in the 1980s, when researchers did not have access to new neurological evidence—but many contemporary researchers still use single arousal instruments or combine the vigor[Mood] and fatigue[Mood] scales to gain a univariate score consistent with outdated theories, perhaps due to their intuitiveness and strong legacy.

This is unfortunate because no reported effect of ICT use on a combined scale of vigor[Mood] and fatigue[Mood] or other univariate scales can be interpreted meaningfully in terms of subjective energy or fatigue. For example, an observed negative relationship between ICT use and a combined vigor[Mood]–fatigue[Mood] construct does not reveal why the relationship is negative. Is it because ICT use is de-energizing, fatiguing or both? Or maybe because the fatiguing effect is stronger than the energizing effect if they run in the opposite direction? Due to the blurred results acquired in this way, in our review we only include those studies that have measured energy and fatigue on separate scales. That is, we only include studies that have either explicitly pursued a bivariate measurement strategy or have used one of the two scales, which still allows us to observe an effect on at least one dimension at a time.

### Energy and Fatigue as Work- and Motivation-related Constructs

Scholars outside the affect sciences who have been interested in work-related and motivation-related energy and fatigue have adopted many baseline assumptions and terms from the underlying affect theories. With the exception of ego-depletion theory (Baumeister 2002) all of them focus on energy and fatigue as subjective experiences. In organizational science and occupational psychology, the experiences of energy and fatigue are termed "(work) vigor" and "emotional exhaustion," respectively (Bakker et al. 2008; Maslach et al. 2001). To highlight their distinction from the affect constructs, we refer to them as vigor[Work] and exhaustion[Work]. However, a challenge is that––at least in organizational science —the two terms are theorized differently than in affect theory. Vigor[Mood] is not the same as vigor[Work] when one studies the details of how these different areas operationalize what they both call by the same name (vigor). Whereas vigor[Mood] captures the momentary feeling of energy without any additional specifics, vigor[Work] captures that same energy feeling but specifically *at work and while working*. The experience of vigor[Work] can also last much longer (it can be permanent), and additionally it encompasses other feelings and attitudes, such as looking forward to going to work (Table 1).

A similar confusion and extension exists around the term "vitality" because vitality is commonly used in motivation psychology (vitality[Mot]), but also it is one of the names used for vigor[Mood]. Judging from its scale, vitality[Mot] is measured in a somewhat similar way to vigor[Mood]; however, the meaning attributed to vitality[Mot] is slightly different in motivation psychology (Deci et al., 2011). First, vitality[Mot] is considered a relatively permanent state rather than a mood that lasts a week or two. Second, next to the core energy feeling, vitality[Mot] also encompasses other feelings and attitudes, such as looking forward to the future (Table 1).

We can summarize these differences between the affect constructs (vigor[Mood] and fatigue[Mood]) and their counterparts from the organizational and motivational sciences (vigor[Work], exhaustion[Work], and vitality[Mot]) to lie on two dimensions: *duration* and *range*. Energy and fatigue are postulated to last much longer in organizational and motivation theories than a simple affect or mood. Emotional exhaustion[Work] is seen as a chronic state rather than a transient experience such as fatigue[Mood].

Furthermore, the conceptual *range* of the energy or fatigue constructs studied in the specialized fields go beyond the pure energetic or fatigued state studied in the affect theories. For example, emotional exhaustion[Work] embraces not only the state of feeling fatigued, but also feeling stressed or strained from work specifically (Maslach et al. 2001). Similarly, vigor[Work] importantly includes mental resilience and willingness to invest effort beyond the general feeling of being energized. Vitality[Mot] includes excitement about the future. With these specificities, the energy and fatigue constructs from the organizational and motivation sciences deviate from the plain and conventional understanding of energetic or fatigued mood. This is not surprising. Scholars working on human energy or fatigue in specific areas naturally refine the constructs to capture the exact phenomenon that applies to their study context. However, this can cause "construct contamination" (DeLuca et al. 2009, p.326).

In terms of bivariate vs. univariate measurement, the organizational sciences are in line with the modern bivariate paradigm.

The construct vigor[Work] is regarded as a distinct counterpart of exhaustion[Work] and only few researchers in the field use a univariate scale that combines vigor[Work] with exhaustion[Work] (Bakker et al. 2008; Demerouti and Bakker 2008).

In the motivational sciences, to our knowledge, there has been no discussion about the bivariate vs. univariate paradigms. The construct vitality[Mot] leans heavily on the affect construct of vigor[Mood] and seems to embrace the problematic univariate conception of energy and fatigue (Ryan and Deci 2008). However, since vitality[Mot] solely measures the energy dimension, we can reliably use it in our review to reveal the effects of ICT use on this single dimension.

### Social Networks-related Fatigue

Recently the IS field has started to look into the fatiguing effects of social networks. IS researchers have used the terms "SNS exhaustion" and "SNS fatigue" to describe these effects. In order to easily distinguish them from the exhaustion[Work] and fatigue[Mood] constructs, we refer to them as exhaustion[SNS] and fatigue[SNS]. While both constructs aim to capture the fatiguing effects of SNS use, they are operationalized differently. Equating them would be misleading because one group of IS researchers borrowed the term "exhaustion" from the organizational sciences and adopted its measures by simply replacing the fatiguing effects of work with the fatiguing effects of SNS use (Ayyagari et al. 2011; Maier et al. 2015). The second group of IS researchers leaned on a mixture of affect theories and borrowed a potpourri of constructs semantically related to "fatigue," such as indifference, boredom and overload. These IS scholars mostly used self-developed scales where they combined fatigue items with items that measure these other constructs (Lee et al. 2016; Zhang et al. 2016).

Another reason why exhaustion[SNS] and fatigue[SNS] should not be equated is again duration. The experience of exhaustion[SNS] is normally conceptualized as longer-lasting because it was directly adopted from exhaustion[Work]. Fatigue[SNS,] on the other hand, is shorter and generally conceptualized as being experienced only around the use of SNS or shortly after.

Finally, IS researchers who study the effects of SNS use on fatigue do not seem to be aware of the bivariate nature of energy and fatigue. This is unfortunate, as any potentially energizing effects from SNS use are thereby automatically factored out.

Table 1 summarizes the four areas and seven categories of energy and fatigue constructs that stem from them. It shows the discussed differences and similarities between the seven construct categories. Considering all of them, we need to ask how much difference the choice of construct makes in the study of ICT use effects on energy and fatigue. In other words, does a choice of energy and/or fatigue construct significantly contribute to the differences in the reported associations? To answer this important question we first review the studies from each area and construct, separately. Before that, we elaborate on our method.

# RESEARCH METHOD

We performed a systematic literature research on the relationship between ICT use and the subjective experience of energy and fatigue. Following Levy and Ellis (2006) and Agogo and Hess (2018), we identified relevant top-ranked peer-reviewed journals in the fields of Information Systems (IS) and Human-Computer Interaction (HCI). To prevent gaps, we further included the leading journals from psychology and organizational behavior, i.e., the research fields from which the energy and fatigue constructs are borrowed.

In a first step, we identified the top 25 journals in the IS and HCI domains and the top five in the external domains using the Scimago Journal and Country Rank database ranks the journals based on the SJR2 indicator (Guerrero-Bote and Moya-Anegón 2012).

In a second step, we searched these journals using the search terms "subjective energy," "mental energy," "psychological energy," "vitality," "vigor," "vigour," "energetic activation," "positive arousal," "fatigue", "exhaustion" OR "depletion." For the journals outside the IS and HCI domains we additionally restricted the search (with the AND operator) to include "computer," "digital," "ICT," "information technology," "laptop," "media," "phone" OR "social network." We used the Web of Science, Ebsco, ProQuest, and Science Direct databases to perform the search within the journals.
The query had no time restrictions and resulted in 9,413 articles (last stand October 2019). After excluding for duplicates, 506 papers remained, which were then scanned for relevance. Inclusion criteria required that

(1) the study is an original empirical research;
(2) at least one measure of subjective energy or subjective fatigue is included in the study;
(3) at least an indirect relationship between technology use and an energy or fatigue measure(s) is reported.

The review excluded articles that combined the subjective energy and fatigue scales, as well as theoretical papers, papers that measured physiological proxies of subjective energy, papers that used performance measures to measure energy or fatigue, as well as papers that measured arousal on a visual or pictorial scale (because these adhere to a univariate paradigm). The screening resulted in 42 papers presented here (Figure 1). For a summary of the articles, see Table A1 in the Appendix.

Based on the measures used for the respective energy or fatigue construct, we categorized the studies in the seven construct categories and four superordinate area categories presented in Table 1. We then reviewed the literature for each area and construct category separately. Note that five papers reported results from two, i.e., three separate studies, which makes the number of reported studies 48 in total.

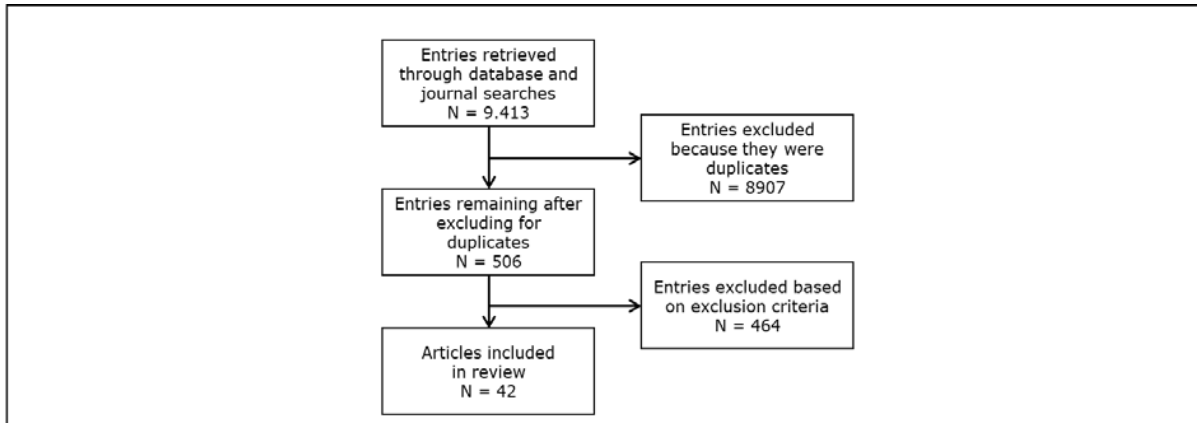

Figure 1. Prism diagram. Flow

Additionally, 12 of the 48 studies considered more than one construct. For example, in their study on cell phone use during work-breaks, Rhee and Kim (2016) examined the effects of such use on vigor$^{Mood}$, fatigue$^{Mood}$, vigor$^{Work}$, as well as exhaustion$^{Work}$. For all these instances (one construct per study), we considered the results separately. Thus, there are four considered instances from the Rhee and Kim (2016) example, one for each construct. In the review, we found 62 such instances, i.e., 14 of the 42 papers are referred to in more than one category (N=9), are considered more than once in a category (N=4), or both (N=1).

Using NVivo12 (QSR International 2018), we coded each instance for the number of reported effects, including zero when the study found no effect of ICT use on the construct in question. It should be noted that many studies found multiple effects of ICT use on the same construct depending on different mediators or moderators. In such cases we coded all of them. Furthermore, when effects were reported, we coded the direction of the effects (i.e., positive vs. negative) and the strength of the effects (i.e., small vs. medium vs. large vs. not reported/cannot be calculated). We derived the strength of effect from the effect sizes reported as standardized beta coefficients from published regressions or structural equation models, the partial eta squares from analysis of variance, as well as Cohen d values from t-tests. Where enough detail was reported, we calculated the effects sizes ourselves from reported mean differences and standard deviations, or from sample sizes and reported statistics (e.g., critical t or F analysis; Faul 2009).

Following Cohen (1988), when we report a "slight or small increase/decrease" in the review, we refer to a small effect size in the relationship between ICT use and energy or fatigue (e.g., ß < .3; Cohen d < .2). When we report a "large or substantial increase/decrease" we refer to a large effect size (e.g., ß >.5; Cohen d < .8). For a detailed summary of the effect sizes in all studies see Table A1 in the Appendix.

If observed, we furthermore denoted all the reported moderators or mediators found to affect the relationship(s) between ICT use and subjective energy or fatigue. Additionally, on a low level we first coded the type of ICTs examined in the study (e.g., e-mail, SNS, cell phone, enterprise system etc.). We then grouped the ICTs into three higher-level categories of ICTs:

(1) conventional ICTs,
(2) organizational ICTs, and
(3) wellness ICTs.

We define "conventional" ICTs when the study in question focused on what is considered a regular or normal use of the Internet; that is, people using the Internet from multiple devices (such as smartphones, laptops, tablets, computers, etc.) for information searches, entertainment, shopping, communication, social networking etc. Organizational ICTs are ICTs and technologies used for work purposes usually in the workplace, such as hardware and software necessary for the performance of work tasks. The use of laptops for remote work was also classified as organizational ICT. Finally, we refer to wellness ICTs when they were explicitly designed to foster people's energy or decrease their fatigue. Examples are virtual reality (VR) technology built to relax and decrease fatigue[Mood], or active video games or applications (known as "exer-games" or "fitness Apps") often designed to increase wellness in general and vigor[Mood] in particular.

Additionally, we coded for the time of ICT use that was examined in the studies (i.e., work hours vs. non-work hours) and the purpose of use (i.e., work-related vs. personal use). Finally, we considered the types of samples and sample sizes used in the studies; the used instruments to measure the construct(s); as well as the used methodology (i.e., cross-sectional design vs. longitudinal design vs. experimental design).

In the following, for each area and category, we summarize the results as beneficial, detrimental or neutral effects of ICT use on energy and/or fatigue. We considered the effects to be beneficial when the studies reported either a positive relationship between ICT use and energy (e.g., increased vigor[Work], vigor[Mood] or vitality[Mot]) or a negative relationship between ICT use and fatigue (e.g. decreased fatigue[Mood], exhaustion[Work], exhaustion[SNS] or fatigue[SNS]). Conversely, the effects are detrimental when ICT use relates to decreased energy and/or increased fatigue. Finally, they are neutral when no relationship was found.

If a study found significant moderators or mediators (even if they are often only driven by a minority), we additionally categorized these results as conditional next to their beneficial, detrimental or neutral classification. When a study reported a direct and an indirect (or moderated) relationship, we counted the two effects separately. Effects dependent on the type of ICT use or type of ICT design were also marked as conditional. For example, if a study reported both a negative and a positive effect because the effect was conditional on personal attitudes, we counted both the positive and the negative effect and additionally categorized these two effects as conditional. On the other hand, if a study found a direct effect of ICT use on fatigue as well as indirect effect through increased information overload, we classified two detrimental effects but only one conditional (on information overload).

Finally, when the study found beneficial and detrimental effects (due to moderators and mediators) we compared the strengths of the beneficial and detrimental effects. We considered one to be larger than the other when the difference was bigger than the threshold for a small effect size ($\beta > .10$; $d > .20$). We performed similar difference tests for studies that considered both dimensions (energy and fatigue), inquiring whether the effects were different on the two dimensions even in the case where both effects were beneficial, i.e., detrimental.

## RESULTS

The extent to which ICT use contributes to exhaustion[Work] was by far the most studied relationship in the reviewed studies (true for N=21 out of 48 studies, or 44% of all studies). We observed medium interest in the effects of ICT use on vigor[Work] (N=10 or 21%) or its influence on affect in general ($N_{vigor}$[Mood]= 8, $N_{fatigue}$[Mood]=6; 17% and 13%, respectively). The fatiguing effect of social networks was studied in 12 studies (25%), twice as many in terms of exhaustion[SNS] (N=8) as in terms of fatigue[SNS] (N=4). Finally, very little interest seems to exist in the effects of ICT use on people's general vitality[Mot], with only five studies (10%) examining this relationship.

Importantly, only a quarter of all the reviewed studies (N=12) used both an energy and a fatigue construct, explicitly aligning with the bivariate view, whereas the rest focused on only one construct. The preferred study method was a cross-sectional survey design (*N*=32, 67%). Only seven studies (15%) used an experimental design and eight (17%) used a longitudinal study design. The median sample size in the reviewed studies was *Mdn*=320 study participants (*M*=517, *SD*=969), ranging from six astronauts in the study with the smallest sample size to 6795 Internet users who watch cat videos in the study with the largest sample size (Table A1, Appendix A). The samples were most often employee samples, (*N*=22, 46%), followed by student samples (*N*=14, 29%) and SNS user samples (*N*=11, 23%).

### *The Effects of ICT Use on the Affective (Mood) States of Energy and Fatigue*

As noted, eight of the 48 reviewed studies looked into the relationship between ICT use and vigor[Mood]. Six of those eight also considered fatigue[Mood], thus explicitly adhering to the bivariate paradigm. The affective state of vigor[Mood] is defined here as the mere feeling of being vigorous, energetic, active, full of pep, and vital—whereas fatigue[Mood] is the feeling of being fatigued, worn-out, exhausted, depleted or tired (McNair et al. 1989; Shacham 1983; Thayer 1986; Ware Jr and Sherbourne 1992).

In this research area there was a pronounced interest in wellness ICTs (see the method section for details). Exactly half (N=4) of the studies that focused on affect constructs studied technologies designed to foster people's energy or decrease fatigue for their effectiveness. Three of the remaining studies looked at conventional ICTs and one at organizational technology. Table 2 summarizes the beneficial, neutral or detrimental ICT effects reported in the eight studies in general, and specifically for conventional, organizational and wellness ICTs. We report on the effects for each group separately because conventional ICTs and organizational ICTs are not developed with the explicit aim of increasing the energy or decreasing the fatigue of their users. The differences between the two and wellness ICTs are therefore particularly interesting with regard to our research question.

As can be seen from Table 2, both conventional and wellness ICTs can have *beneficial effects* on mood, i.e., they can increase vigor[Mood] as well as reduce fatigue[Mood]: Two out of the five beneficial effects that we found on vigor[Mood] and one out of the two we found on fatigue[Mood] stemmed from wellness ICTs. The remaining beneficial effects were due to conventional ICTs (3 and 1, respectively).

However, when it comes to *detrimental ICT effects*, conventional ICTs seem more likely to cause them: Only in one case did a wellness ICT report detrimental effects on mood, compared to four such results reported for conventional ICTs. Note that the number of studies considering positive ICTs is similar to those considering conventional ICTs, so this comparison is justified.

What we also see from Table 2 is that at least in terms of numbers of observed effects, conventional ICT use leads to just as many beneficial effects as detrimental ones, both on vigor[Mood] and fatigue[Mood]. Hence, from a very high-level perspective, the affective states vigor[Mood] and fatigue[Mood] can be influenced in both directions through the conventional use of the Internet. As we delve into the details of these effects, though, we find important nuances to this picture.

**Table 2 . Number of reported results that revealed a beneficial, neutral or detrimental effect of ICT use on vigor[Mood] (upper half) and fatigue[Mood] (lower half)**

| ICT use effects on **Vigor[Mood]** | | | | | | | | | | | | |
|---|---|---|---|---|---|---|---|---|---|---|---|---|
| Type of effect: | **ICT use is beneficial** (increases vigor[Mood]) | | | **ICT use is neutral** (no effect) | | | **ICT use is detrimental** (decreases vigor[Mood]) | | | **Conditional effects/Total effects** | | |
| **Type of ICT:** | Con. | Org. | Well. | Con. | Org. | Well. | Con. | Org. | Well. | Con. | Org. | Well. |
| Number of effects: $N_{Con}=3, N_{Org}=1, N_{Well}=4$ | **3** | 0 | 2 | **2** | 1 | 1 | **3** | 0 | 1 | 8/8 | 1/1 | 0/4 |
| Total effects (N=8) | **5** | | | **4** | | | **4** | | | **9/13** | | |
| ICT use effects on **Fatigue[Mood]** | | | | | | | | | | | | |
| Type of effect: | **ICT use is beneficial** (decreases fatigue[Mood]) | | | **ICT use is neutral** (no effect) | | | **ICT use is detrimental** (increases fatigue[Mood]) | | | **Conditional effects/Total effects** | | |
| **Type of ICT:** | Con. | Org. | Well. | Con. | Org. | Well. | Con. | Org. | Well. | Con. | Org. | Well. |
| Number of effects: ($N_{Con}=2, N_{Org}=1, N_{Well}=3$) | **1** | 0 | 1 | **0** | 0 | 3 | **1** | 1 | 0 | 2/2 | 1/1 | 0/4 |
| Total effects (N=6) | **2** | | | **3** | | | **2** | | | **3/7** | | |
| Note: Con. = conventional ICTs, Org. = organizational ICTs, Well. = positive ICTs; N = number of studies from which the results (effects) stem. | | | | | | | | | | | | |

## Beneficial Effects of ICT Use for Mood

When we look into the details of the reported results, we see that both fitness games and mood-inducing technologies can successfully increase the experience of vigor[Mood], but only VR, and not fitness games, was reported to decrease fatigue[Mood] (Herrero et al. 2014; Huang et al. 2017; Lee et al. 2017). For conventional ICTs, we find beneficial effects at moments where Internet use is self-determined or after viewing entertaining Internet content that makes people happy. For example, Myrick (2015) has shown that people retrospectively report slightly higher vigor[Mood] and substantially lower fatigue[Mood] after watching online cat videos in comparison to how they felt before watching them. This study suggests that users believe using ICTs for entertainment can decrease their fatigue and even increase their energy. If so, they might consciously turn to this type of online entertainment when they feel fatigued or low on energy.

In a diary study with heavy Internet users, Quinones and Griffiths (2017) have supported this assumption and shown that employees are more likely to use the Internet when their workday is demanding and energy-draining. This happens both during working hours (a behavior also known as "cyberloafing") and in the evening after work. This increased Internet use then indeed increased employees' vigor$^{Mood}$ showing that ICT use can be an efficient energizer.

**Detrimental Effects of ICT Use for Mood**

The caveat is that such beneficial effects of intense Internet use for private purposes at work (cyberloafing) and at home seem only true for people who have a less problematic relationship with the Internet to begin with; that is, users who embrace the Internet uncritically and do not try to control their excessive use.

In contrast, Internet use was shown to be *detrimental* for people who try to resist and control their use but fail to do so. For this latter group of otherwise ordinary users, intense Internet use at work or at home consistently decreased their vigor$^{Mood}$ (Quinones and Griffiths 2017). What is more, the detrimental effects of ICT use on users who try to control their use seem to be stronger than the beneficial effect of ICT use for users who do not seem to view their excessive Internet use as a problem. There seems to exist a threshold of Internet use *appraisal* that influences whether ICT use has beneficial or detrimental effects on vigor$^{Mood}$.

Other studies have shown that subsequent negative attitudes (such as guilt), which often follow use-control failures, contribute to a decreased energetic mood (Myrick 2015; Reinecke et al. 2014). The use-control failure effect has been repeatedly demonstrated on mood for multiple technologies such as games, TV, the Internet, and specifically SNS (Du et al. 2018; Kelley and Gruber 2010). According to the self-determination theory, a failure to control one's ICT use is a hallmark of lack of autonomy, and comes with feelings of being pressured, coerced or seduced to use ICTs (Deci and Ryan 2000). More precisely, it is the feeling of being controlled by the ICTs or by one's own urge to use ICTs. This personal *lack of autonomy* and the *inner conflict* that it brings seem responsible for the drain of one's energy (Ohly and Latour 2014).

In line with this autonomy argument, in a study about four different levels of airplane automation and pilot experience, Hancock (2007 has convincingly shown that the more autonomy a pilot has in controlling the airplane the less fatigued they feel after flying it. The same study found no effect on vigor$^{Mood}$, further showing that the effect of technology on UX can be different on the two dimensions, in line with the bivariate paradigm. The demonstrated stronger effects of ICT use on fatigue$^{Mood}$ than on vigor$^{Mood}$, add to the already reported stronger effects on fatigue$^{Mood}$ found by Myrick (2015).

Rhee and colleagues (2016) further demonstrated a stronger effect of ICT use on fatigue$^{Mood}$ than on vigor$^{Mood}$. In their study of cell phone use during work breaks they compared the effects of such use with conventional work breaks (e.g., employees have lunch, talk to their colleagues or take a walk). They found that whereas cell phone work breaks do not alter the feeling of

vigor[Mood] in comparison to conventional breaks, taking breaks on a cell phone significantly increases fatigue[Mood] (Rhee and Kim 2016). The reason for the increased fatigue[Mood] was the fact that cell phone use prevents employees from detaching from work and properly recuperating in the same way as they would in less stimulating environments (Berman et al. 2008; Berto 2005; Herzog et al. 2003).

In sum, the studies on the effects of ICT use on mood show that ICTs can have a beneficial effect for both vigor[Mood] as well as fatigue[Mood], with positive ICTs being especially beneficial for vigor[Mood]. The effects on both dimensions can be equally detrimental, especially in the case of excessive use of conventional ICTs, accompanied with an inability to control usage. Here, the strength of the detrimental effects was also stronger than the beneficial. Finally, the effects on fatigue[Mood] were also often stronger than the effects on vigor[Mood], regardless of whether the effect was beneficial or detrimental.

### The Effects of ICT Use on Energy for Work and Fatigue from Work

As many as half of all reviewed studies (N=24, 50%) used organizational constructs, i.e., examined the relationship between ICT use and vigor[Work] (N=3), exhaustion[Work] (N=14), or both (N=7). Vigor[Work] in the organizational sciences is defined as "high levels of energy *while working,*" but this can include mental resilience and persistence at work, as well as inclination towards the work (Schaufeli et al. 2002, p. 74). Emotional exhaustion[Work], on the other hand, is conceptualized as the feeling of chronic fatigue caused by one's work, i.e., feeling used up and emotionally drained by one's work (Maslach et al. 2001; Moore 2000; Schaufeli et al. 1996). Both constructs relate only to the working population.

Importantly, five reviewed studies adopted the vigor[Work] or exhaustion[Work] constructs to mean vigor for school or university work, i.e., exhaustion from school or university work. Since this type of work is fundamentally different from typical employee's work, we present the results of these studies separately. The studies on the effects of ICT use on vigor[Work] and exhaustion[Work] focused mostly on *work-related* ICT use, often regardless of the type of ICT. In these studies, the time and place of work was an important focus, whereas the ICT could be conventional ICT (e.g., smartphones); generic technology (e.g., word processors and spreadsheets); or more specific organizational ICTs (e.g., enterprise systems, knowledge and safety management systems, database technologies, etc.).

Specifically, the effects of *work-related* ICT use *after work hours* has been subject to considerable investigation (N=11), followed by the use of new information systems and its impact on employees at work (N=7). Research has also focused on remote work ('teleworking', N=5), and ICT use for private purposes among students (N=4), whereas less interest was shown in ICT use for private purposes among employees during official office hours (a behavior known as "cyberloafing" and "cyberslacking"; N=2). Table 3 presents the effects of ICT use on vigor[Work] and exhaustion[Work] in an employment and student context, as well as overall.

**Table 3. Number of reported results that revealed a beneficial, neutral or detrimental effect of ICT use on vigor$^{Work}$ and exhaustion$^{Work}$**

| | Vigor$^{Work}$ (N=10) | | | | | | | |
|---|---|---|---|---|---|---|---|---|
| Type of effect: | ICT use is beneficial (increases vigor$^{Work}$) | | ICT use is neutral (no effect) | | ICT use is detrimental (decreases vigor$^{Work}$) | | Conditional effects/Total effects | |
| $N_{Emp}$=5; $N_{St}$=5 | Empl. | Stud. | Empl. | Stud. | Empl. | Stud. | Empl. | Stud. |
| Number of effects | 5 | 10 | 5 | 9 | 6 | 6 | 14/16 | 25/25 |
| Total effects | 15 | | 14 | | 12 | | 39/41 | |
| | Exhaustion$^{Work}$ (N=21) | | | | | | | |
| Type of effect: | ICT use is beneficial (decreases exhaustion$^{Work}$) | | ICT use is neutral (no effect) | | ICT use is detrimental (increases exhaustion$^{Work}$) | | Conditional effects/Total effects | |
| $N_{Emp}$=18, $N_{St}$=3 | Empl. | Stud. | Empl. | Stud. | Empl. | Stud. | Empl. | Stud. |
| Number of effects | 18 | 1 | 13 | 9 | 27 | 6 | 52/58 | 16/16 |
| Total effects | 19 | | 22 | | 33 | | 68/74 | |

Note. Empl. = studies focusing on employees' vigor$^{Work}$ and exhaustion$^{Work}$, Stud. = studies focusing on students' vigor for and exhaustion from school or university work; N = number of studies from which the results (effects) stem.

At first view, it is evident that detrimental effects are more frequently found for exhaustion$^{Work}$ than for vigor$^{Work}$. Whereas less than a third (12, or 29%) of all reported effects on vigor$^{Work}$ showed ICT use can decrease it and a bit more than a third (15, or 37%) showed that ICT use can increase vigor$^{Work}$, for exhaustion$^{Work}$ it was reversed. Almost a half of all reported effects (33, or 45%) showed ICT use increases exhaustion from work and only a quarter (19, or 26%) showed a decrease.

Notably, the majority of the beneficial effects for vigor$^{Work}$ were reported for students, i.e., for vigor for students' type of work. Here the beneficial to detrimental effects ratio (10:6) is more positive than for employees (5:6). For exhaustion$^{Work}$, the picture is again revered. The majority of beneficial effects come from employees and for them the beneficial to detrimental effects ratio (18:27) is less negative than for students (1:6), but is notably large nonetheless.

Finally, almost all of the reported effects are conditional, which means they are dependent on attitudes, personality traits, the type of ICT use, etc. We now turn to the details of these effects in order to better understand the contexts in which ICT use is beneficial and those in which it is detrimental.

**Beneficial Effects of ICT Use on Work-related Energy and Fatigue**

The most common context in which ICT use increases vigor$^{Work}$ is (surprisingly) work-related, after-work hours ICT use, such as answering work e-mail or using the Internet for work-purposes at home (Table 4). This type of use was related to higher vigor$^{Work}$ in all seven studies that looked at this context, which found a total of eleven beneficial effects (Table 4).

The majority of these effects are only true for employees and students who voluntarily engage in after-hour work-related activities (van Zoonen and Rice 2017), prefer to respond immediately to colleagues (Ragsdale and Hoover 2016), and who can themselves control the time of their ICT use (Llorens et al. 2007). In students, vigor is further related to self-directed learning, i.e., knowledge-oriented Internet use (Hietajärvi et al. 2019; Rashid and Asghar 2016).

Taken together the results suggest that the construct of employee autonomy plays a role here: vigor$^{Work}$ is increased when ICTs are used autonomously (Llorens et al. 2007; Ragsdale and Hoover 2016; Rashid and Asghar 2016) or when their use increases the perception of autonomy (van Zoonen and Rice 2017). This observation has support in the context of teleworking too, where increased autonomy is the largest predictor through which ICT use increases vigor$^{Work}$, i.e., the only effect of medium size (Sardeshmukh et al. 2012).

| Table 4. Number of found effects on organizational constructs depending on context | | | | | | | | | | | | | | | | |
|---|---|---|---|---|---|---|---|---|---|---|---|---|---|---|---|---|
| Time of use | Work hours | | | | | | | | After-work hours | | | | | | | |
| Type of use | Work-related use | | | | Personal use | | | | Work-related Use | | | | Personal Use | | | |
| Type of effect | + | 0 | - | C | + | 0 | - | C | + | 0 | - | C | + | 0 | - | C |
| Vigor$^{Work}$ | 3 | 3 | 3 | 8 | 1 | 0 | 0 | 1 | 11 | 4 | 2 | 16 | 5 | 9 | 6 | 20 |
| Exhaustion$^{Work}$ | 14 | 10 | 13 | 34 | 0 | 0 | 4 | 3 | 4 | 7 | 8 | 17 | 1 | 6 | 6 | 13 |

Note. Type of effect: "+" = ICT use is beneficial, i.e., it increases vigor$^{Work}$ or decreases exhaustion$^{Work}$; "0" = ICT use is neutral, i.e. no effect was found; "- "= ICT use is detrimental, i.e. it decreases vigor$^{Work}$ or increases exhaustion$^{Work}$. "C" = conditional effects.
The effects of two studies are presented in two contexts because they spanned across them (Llorens et al. 2007; Rashid and Asghar 2016).
The effects of the studies on teleworking are presented in the work-hours section, even though we are aware that the nature of telework can blur the line between work-hours and after-work hours.

It is important to note, however, that a reverse causation between vigor$^{Work}$, autonomy and ICT use is also likely. It is plausible that employees who feel energetic about their jobs are consequently more likely to both voluntarily use ICTs for work purposes in their private time as well as use them more often. Unfortunately, only one study in a student setting tested for a reverse relationship and confirmed reciprocal causations: Autonomous ICT use increases student vigor and vigor in turn increases autonomous ICT use (Llorens et al. 2007).

Beneficial effects of ICT use are also evident on the fatigue dimension, that is, when ICT use decreases exhaustion$^{Work}$. Interestingly, two thirds (N=14) of all studies reported decreased exhaustion caused by ICT use at least for some employees or some contexts, even if that constituted only a quarter of all beneficial effects (Table 3). Decreased exhaustion in relation to ICT use also happens when ICT use is self-determined, that is, when it supports employees' autonomy (Bala and Bhagwatwar 2018; Piszczek 2017; Sardeshmukh et al. 2012; van Zoonen and Rice 2017; Xie et al. 2018).

Employees' expectations (Piszczek 2017), preferences (Piszczek 2017; Ragsdale and Hoover 2016; Xie et al. 2018), values (Hennington et al. 2011) or personality (Gaudioso et al. 2017) also play a role, and most of them are again related to autonomy. For example, those employees who prefer to integrate their work and personal lives preferences (Piszczek 2017; Xie et al. 2018) or prefer to respond immediately (Ragsdale and Hoover 2016), show lower exhaustion[Work] related to work-related ICT use after work-hours. Employees who can control their work-life boundary (Piszczek 2017), and regulate when to respond to work-related messages also show lower exhaustion[Work] (van Zoonen and Rice 2017).

Other variables not related to autonomy were also shown to lower exhaustion[Work]. For example, when work-related after-hours ICT use allows employees to finish open tasks, this can slightly decrease exhaustion[Work] (Chen and Karahanna 2018). Remote workers can also profit with regards to exhaustion[Work] because telework slightly decreases time pressure (Sardeshmukh et al. 2012) and helps lower the stress arising from interpersonal conflict in the workplace (Sardeshmukh et al. 2012; Windeler et al. 2017). This makes remote work a desirable context in which ICTs can help against exhaustion[Work]. When it comes to introduction of new ICTs in the workplace, computer training can help lower exhaustion[Work,] too, but only if the training is successful (i.e., employees use the system features taught in training (Bala and Bhagwatwar 2018), and when the training increases employees' ICT self-efficacy (Salanova et al. 2000).

A note of caution when interpreting the beneficial effects of work-related ICT use on exhaustion[Work] is necessary here as well, however, because a reverse causation is possible. Unfortunately, only one paper tested for such a relationship albeit in two separate studies for two very different systems and organizations. It showed that employees are less likely to use newly introduced organizational ICTs when they feel exhausted by their work to start with

(Bala and Bhagwatwar 2018). In this case, it was not that ICT use decreased exhaustion[Work] but exhausted employees were less likely (and probably less willing) to frequently use the newly introduced organizational ICTs, a conclusion that might be missed with a cross-sectional design.

**Detrimental Effects of ICT Use on Work-related Energy and Fatigue**

As the beneficial effects of ICT use on vigor[Work] and exhaustion[Work] are mostly conditional on employees' attitudes, expectations, personality or values, we also see many detrimental effects of ICT use on vigor[Work] and exhaustion[Work] when these conditions are not met.

For example, just as vigor[Work] is higher among those employees who are "attached" to their phones and prefer to respond immediately to work-related messages during after-work hours, the opposite is true for those who feel no such attachment, i.e., have no such preference. For them, work-related cell phone use during after-work hours drains their energy for work (Ragsdale and Hoover 2016). What is more, the beneficial effect of ICT use on autonomy and thereby on vigor[Work], is only present for those employees who do not immediately respond to work messages and who are not always available for their colleagues' requests (van Zoonen and Rice 2017). For remote workers, vigor[Work] can be drained in cases of no technical support and technical failures (Chen 2017), but also because telework slightly decreases collegial support and feedback (Sardeshmukh et al. 2012).

The detrimental effects of ICT use are even more pronounced for exhaustion[Work]. Seventeen out of the 21 reviewed studies show that ICT use is related to increased exhaustion[Work], at least among some group of employees. And just as with the effects found on vigor[Mood] and fatigue[Mood], the detrimental effects found on exhaustion[Work] were stronger than the beneficial. Only one of the thirteen studies that found beneficial and detrimental effects on exhaustion[Work] showed stronger beneficial effects, whereas seven showed stronger detrimental effects. Moreover, even when the strength of effects was similar, there were usually more participants for whom the effects of ICT use were detrimental than beneficial, or there were overall more detrimental effects. Conversely, the seven studies that found both beneficial and detrimental effects on vigor[Work], showed mostly similar number of effects with similar strength. Similarly, the few studies that looked at both dimensions did not find big discrepancies in strength, but they did find more detrimental effects on exhaustion[Work] than on vigor[Work], making the overall detrimental effect on exhaustion[Work] more significant.

Looking at the details for the detrimental effects, three main patterns through which ICT use increases employees' exhaustion[Work] emerge, namely,

(1) the *overload* and *pressure* that arises from the ubiquitous nature of ICTs, from the interruptions that are its consequence, and from the pace of change that they introduce;

(2) the *conflicts* that result from the use, overload and created pressure;

(3) *the lack of autonomy* instigated by employer expectations for mandatory ICT use.

First, ICTs make employees available for interruptions anywhere and at any time during office hours (Ayyagari et al. 2011; Gaudioso et al. 2017), but also at home and during after-work hours (Piszczek 2017; Ragsdale and Hoover 2016; Xie et al. 2018). ICT ubiquity has shifted expectations of the availability of employees outside of working hours and has blurred the so-called work-life boundary (Mazmanian et al. 2013; Piszczek 2017). As a result, employees have no private sphere left that allows them to detach, unwind or recover from the workday, a condition termed "techno-invasion" (Gaudioso et al. 2017). ICTs also increase complexity and the pace of work (Ayyagari et al. 2011; Chen and Karahanna 2018; Gaudioso et al. 2017). All these changes create overload, such as *work overload, information overload, interruption overload, e-mail overload,* or generally, techno-overload—as well as pressure, such as *time pressure* and *work pressure.* Addiction to ICTs and their excessive use can additionally increase overload, simply because ICTs "lure" employees to cyberloaf and thus procrastinate, which causes work to pile up (Aghaz and Sheikh 2016).

In all these cases, the employees are left feeling more exhausted by their jobs because of ICT use. Chen and Karahanna (2018) have further shown that these work-related interruptions and the subsequent overload leave the employees emotionally exhausted not only from their work, but also from the demands of their personal lives.

Ironically, employees most often cyberloaf when they try to recover energy—even though this behavior actually increases their exhaustion[Work] (Aghaz and Sheikh 2016). Even taking work-breaks on a cell phone increases exhaustion[Work] in comparison to conventional work-breaks because, as we have shown above, ICT use during work-breaks increases fatigue[Mood], and fatigue[Mood] after breaks in turn increases exhaustion[Work] ((Rhee and Kim 2016).

Second, ICT use and ICT interruptions have been shown to cause many conflicts, most notably role conflicts and work-family (work-home) conflicts (Ayyagari et al. 2011; Hennington et al. 2011; Ragsdale and Hoover 2016; Zheng and Lee 2016). These conflicts are one of the strongest predictors of exhaustion[Work]. Whereas the main, direct effect of work-related ICT use outside of working hours on exhaustion[Work] is usually non-significant (Piszczek 2017) or small (ß =.20 to ß =.27; Ragsdale and Hoover 2016; Xie et al. 2018), the effect it has indirectly through conflict is much larger (Ayyagari et al. 2011; Ragsdale and Hoover 2016; Zheng and Lee 2016).

Third, next to the overload, pressure, and conflict, a consistent pattern through which ICT use has been linked to employees' increased exhaustion[Work] is the lack of autonomy that sometimes comes with digitalization of the workplace. The introduction of new organizational systems and the mandatory training, adaptation and use that they require have been found to exhaust employees, especially after inadequate training (Bala and Bhagwatwar 2018; Salanova et al. 2000). Again, it is the mismatch between the introduction of new ICTs and employees' expectations, preferences, and values that play a significant role. Namely, the digitalization of the workplace exhausts especially those employees who see the introduction of the ICTs and their mandatory use as *controlling* or not compatible with their personal values (Bala and Bhagwatwar 2018; Hennington et al. 2011).

Other negative effects of ICT use conditional on the employees' personality have also been demonstrated. For example, Reinke and Chamorro-Premuzic (2014 have shown that the relationship between perceived e-mail overload and exhaustion[Work] has more to do with an employee's personality than the actual number of received e-mails. Some employees have worse coping strategies with stress caused by technology (Gaudioso et al. 2017); others prefer to segregate their work from their personal life and do not want to be available to colleagues outside of working hours (Ragsdale and Hoover 2016; van Zoonen and Rice 2017; Xie et al. 2018). All these employees feel emotionally exhausted from work-related ICT use during and after working hours (Piszczek 2017; Ragsdale and Hoover 2016). Thus, it seems that the readiness to work with organizational ICTs as well as the feeling of being in control and self-determined (autonomous) in work-related ICT use are important buffers against exhaustion[Work].

That said, Xie et al. (2018) have shown that work-related after-hours ICT use continues to predict exhaustion[Work] even after controlling for autonomy or work-integration preferences. This means that ICT use can exhaust even when employees engage in it voluntarily and prefer to work outside of working hours, although to a much lesser extent. What is more, the majority of employees want to separate their work and personal lives and want to be in control of where, when and how they use work-related ICTs (Bala and Bhagwatwar 2018; Hennington et al. 2011; Xie et al. 2018).

Finally, looking at the different ICTs reported in the studies, we find that not all technologies are equal in the extent to which they cause exhaustion[Work]. For example, answering e-mails outside of office hours is not as detrimental as using the phone or messaging (Chen and Karahanna 2018). SNS use, on the other hand, has an overall negative effect both for employees (Aghaz and Sheikh 2016; van Zoonen and Rice 2017) and for students (Hietajärvi et al. 2019).

In sum, ICT use in the work-related context has more beneficial effects for vigor[Work] than for exhaustion[Work]. The most general condition under which ICT use helps increase vigor[Work] and decrease exhaustion[Work] is autonomy and vice versa. In the case of exhaustion[Work], overload, pressure and conflict caused by ICT use can further substantially increase exhaustion[Work]. Interestingly, most contexts that were shown to invigorate were the same contexts related to increased exhaustion, suggesting that ICT use can have very different effects on the two dimensions. For example, using ICTs outside of working hours or during work-breaks is very often related to high vigor[Work], but at the same time, this type of use is associated with higher exhaustion[Work]. Similarly, while employees cyberloaf in order to increase their energy, they leave themselves vulnerable to increased exhaustion[Work], too (Aghaz and Sheikh 2016). These potentially parallel invigorating and draining processes might be amplified by ICT use. For example, in the study by Rhee and Kim (2016), there was no correlation between the experiences of exhaustion[Work] and vigor[Work] among the employees who use their smartphones on work-breaks, whereas there was medium sized negative correlation for those who take "conventional" work-breaks. In other words, for those who use conventional breaks the intuitive is true: feeling more vigorous corresponded to feeling less exhausted. For employees who use smartphones on their work-breaks the experiences of exhaustion[Work] and vigor [Work] were completely unrelated.

Unfortunately, only seven (15%) of the reviewed studies in this area explicitly adhered to the bivariate paradigm and showed this trend, i.e., examined the effects of ICT use on both vigor[Work] and exhaustion[Work] in parallel to shed light on this potential paradox.

## The Effects of ICT Use on Vitality

The construct "vitality" (or *subjective vitality*[Mot]), as defined in self-determination theory (SDT), is a feeling of aliveness and vigor accompanied by a long-lasting state of calm energy (Ryan and Deci 2008). SDT postulates that vitality[Mot] is a direct consequence of the (not necessarily conscious) satisfaction of the three basic psychological needs, that is, the need for self-determination, i.e., autonomy; the need for efficacy, i.e., competence; and the need for relatedness, i.e., belongingness.

Only five of the 42 reviewed studies investigated subjective vitality[Mot], and all of them focused exclusively on ICT use for private purposes. Four studies looked at conventional ICTs, i.e., the Internet or Social Networking Sites, and one looked at fitness apps, i.e., wellness ICTs.

Table 5 summarizes the reported overall relationship between ICT use and vitality[Mot] to suggest that ICT use and especially conventional ICT use can be rather detrimental for long-term vitality. Looking at the details, two studies found a slight negative association between vitality[Mot] and excessive Internet use ($\beta = -.13$, Akın 2012) and excessive SNS use specifically ($\beta = -.24$; Satici and Uysal 2015). This is in line with SDT theory, which posits that it is the feeling of being pressured, coerced or even seduced to succumb to one's own urge to use ICTs that causes this effect (Ohly and Latour 2014). In other words, it is the diminished autonomy over the ICT use and the inner conflicts that come with it that seem to deplete vitality[Mot].

| Table 5. Number of reported results that revealed a beneficial, neutral or detrimental effect of ICT use on vitality$^{Mot}$ | | | | | | | |
|---|---|---|---|---|---|---|---|
| Examined construct | ICT use is beneficial (increases vitality$^{Mot}$) | | ICT use is neutral (no effect) | | ICT use is detrimental (decreases vitality$^{Mot}$) | | Conditional effects | |
| Type of ICT/ Construct: | Con. | Well. | Con. | Well. | Con- | Well | Conv. | Well, |
| | 0 | 6 | 2 | 5 | 2 | 4 | 7/4 | 15/15 |
| Vitality$^{Mot}$ (N=5) | 6 | | 7 | | 6 | | 19/19 | |
| Note: Con. = conventional ICTs; Pos. = positive ICTs; N = number of studies from which the results stem. | | | | | | | | |

The other two studies focused on the interplay between some ICT features and users' motivation and personality. In the case of fitness apps, James et al. (2019) found that users who are intrinsically motivated to do sports, as well as those who lack motivation to exercise, could profit from the data management and social features of these apps. The same features were found to thwart vitality$^{Mot}$ for extrinsically motivated users, however, emphasizing the nuanced effects that ICTs features can have on human energy. On the other hand, the option to strategically or authentically present oneself on SNS, which can be theorized as a potential beneficial feature of SNS, does not seem to impact vitality$^{Mot}$, regardless of personality (Jang et al. 2018).

In sum, the few studies that focused on vitality$^{Mot}$ reiterate the detrimental effects of excessive Internet and SNS use for human energy demonstrated for the mood and work-related constructs.

### SNS Exhaustion and SNS Fatigue

Seen the repeated detrimental effects of SNS reported on all prior constructs, it is no wonder that two new fatigue constructs have emerged in the IS and HCI literature: exhaustion$^{SNS}$ and fatigue$^{SNS}$ (Table 1). Beyond "traditional" SNS such as Facebook, Qzone, Twitter or Instagram, the concept of SNS fatigue has also been applied to mobile instant messengers (Shin and Shin 2016). To our knowledge, no construct of vigor$^{SNS}$ has been considered in the literature, so the focus in this area of research is exclusively on the fatigue dimension. Specifically, most of this research looks for the causes of the established fatiguing effects of SNS use.

The two constructs of exhaustion$^{SNS}$ and fatigue$^{SNS}$ are defined as reactions to SNS use, which demonstrate themselves in feelings of fatigue and exhaustion (Bright et al. 2015; Lee et al. 2016; Luqman et al. 2017). Whereas exhaustion$^{SNS}$ is a narrower but longer experience of feeling drained by SNS use, fatigue$^{SNS}$ is conceptualized as a broader experience that next to the experience of fatigue from SNS use also considers the rise of emotional states such as indifference to or boredom from SNS use (Table 1). For example, whereas the items "I feel drained," and "I feel tired from my Facebook activities" measure both constructs (Lee et al. 2016; Maier et al. 2015; Zhang et al. 2016), "When I use QZone, I feel bored" measures fatigue$^{SNS}$ only (Zhang et al. 2016).

Of the 39 effects reported in the 12 studies analyzing exhaustion$^{SNS}$ and fatigue$^{SNS}$, only two were beneficial and the rest detrimental, confirming the detrimental potential of SNS use already established by the studies on mood, exhaustion$^{Work}$ and vitality$^{Mot}$ (Table 6).

Notably, five studies even reported high exhaustion[SNS] or fatigue[SNS] in their samples overall (mean values were over the midpoint of the scale). No matter the average experience of exhaustion[SNS] or fatigue[SNS] within the study participants, all but one of the reported effects were conditional (38 of 39, Table 6). This means that whether one experiences exhaustion[SNS] or fatigue[SNS] depends on other factors. We thus turn to these factors and conditions under which SNS use leads to exhaustion[SNS] and fatigue[SNS].

The most prominent mediator through which SNS use contributes to both exhaustion[SNS] and fatigue[SNS] is perceived overload in its various forms. Seven out of 12 studies reported at least one type of overload as an indirect condition for exhaustion[SNS] (N=4) or fatigue[SNS] (N=3). These studies distinguish between information overload, social overload, communication overload and system feature overload.

**Table 6. Number of reported results that revealed a beneficial, neutral or detrimental effect of ICT use on exhaustion[SNS] and fatigue[SNS]**

| Examined construct (Number of studies) | ICT use is beneficial (decreases exhaustion[SNS] or fatigue[SNS]) | ICT use is neutral (no effect) | ICT use is detrimental (increases exhaustion[SNS] or fatigue[SNS]) | Conditional effects |
|---|---|---|---|---|
| Exhaustion[SNS] (N=8) | 1 | 3 | 20 | 23/24 |
| Fatigue[SNS] (N=6) | 1 | 2 | 12 | 15/15 |
| Overall effects for IS constructs | 2 | 5 | 32 | 38/39 |
| Note: N = number of the studies from which the results (effects) stem. | | | | |

*Information overload* is the feeling of having to deal with an overwhelming amount of information on social networks, i.e., feeling distracted by this excessive amount of available information (Cao and Sun 2018; Gao et al. 2018; Lee et al. 2016; Luqman et al. 2017; Zhang et al. 2016). *Social overload* is the feeling of excessively dealing with friends' problems, paying too much attention to their posts and caring too much about them (Cao and Sun 2018; Lo 2019; Maier et al. 2015; Zhang et al. 2016).
*Communication overload*, on the other hand is feeling overloaded by social media communication, i.e., receiving too many notifications from SNS and consequently having to respond to too many messages (Cao and Sun 2018; Lee et al. 2016; Shin and Shin 2016). All three types of overload have been repeatedly related to exhaustion[SNS] (Cao and Sun 2018; Gao et al. 2018; Lee et al. 2016; Lo 2019; Luqman et al. 2017; Maier et al. 2015; Shin and Shin 2016; Zhang et al. 2016), as well as fatigue[SNS] (Lee et al. 2016; Zhang et al. 2016). Whereas the impact of information overload is constantly small, social and communication overload, which are specific to SNSs, tend to be stronger predictors, as both medium and large effects were reported ($\beta_{info}$ = .20 to $\beta_{info}$ = .26; $\beta_{social}$ = .24 to $\beta_{social}$ =.62; Cao and Sun 2018; Gao et al. 2018; Lee et al. 2016; Lo 2019; Luqman et al. 2017; Maier et al. 2015; Zhang et al. 2016).

*System feature overload*, which refers to distraction caused by too many complex, unnecessary and poorly designed features, was only studied for fatigue[SNS] and was found to slightly increase it (Lee et al. 2016; Zhang et al. 2016). In sum, when SNS users feel any of the various types of overload—but especially communication and social overload—they are more likely to experience exhaustion[SNS] and fatigue[SNS].

Just having an SNS app installed on a smartphone with a permanent Internet connection is an additional route to exhaustion[SNS] and fatigue[SNS]. Ubiquitous connectivity increases the likelihood of exhaustion[SNS] not only directly, but also through increasing the perceived overload (Gao et al. 2018). Overload also increases with each added SNS friend (Maier et al. 2015) and each feature change on the SNS, coupled with the complexity of these features (Lee et al. 2016).

Other reported predictors of exhaustion[SNS] and fatigue[SNS] are excessive use and conflict. Initial social, informational and hedonic reasons for SNS use can prompt excessive use (Luqman et al. 2017), and excessive use was repeatedly linked with exhaustion[SNS] (Cao and Sun 2018; Lo 2019; Luqman et al. 2017; Maier et al. 2015; Zheng and Lee 2016). Excessive use is especially problematic if one develops emotional and cognitive preoccupation with SNS use, typical for addictive behavior (Aboujaoude 2010). Excessive use is also problematic because it causes conflict; specifically, it can lead to technology-personal, technology-family and technology-work conflicts (Zheng and Lee 2016). Considering the reported effects so far, it should come as no surprise that excessive SNS use—and all the different conflicts that arise from such use—directly and indirectly drive exhaustion[SNS], often to a large extent (ß=.36 to ß=.60; Cao et al. 2018; Zheng and Lee 2016).

Additionally, SNSs create exhaustion[SNS] through the increased possibility of social comparison and the shame that can come with it (Lim and Yang 2015). Other mechanisms through which SNS use increases fatigue[SNS] include the ambiguous nature of the information posted on the platforms (Lee et al. 2016), as well as privacy concerns that frequently arise on these platforms (Bright et al. 2015).

A rare positive side of SNS use that can decrease exhaustion[SNS] and fatigue[SNS] was shown for certain personal characteristics; namely, the social support that users can get on such platforms decreases their exhaustion[SNS] if the users are lonely, but emotionally stable (Lo 2019). Additionally, perceived self-efficacy in SNS use decreases fatigue[SNS] (Bright et al. 2015).

Finally, a potential reverse causation between SNS fatigue and SNS use was also reported. More than half of the studies in this domain found that both exhaustion[SNS] and especially fatigue[SNS] are related to discontinuous usage intentions (Cao and Sun 2018; Gao et al. 2018; Lo 2019; Luqman et al. 2017; Maier et al. 2015; Shin and Shin 2016; Zhang et al. 2016).

In sum, three main drivers already reported for the established constructs of energy and fatigue from the affective, organizational and motivational literature, have also been found to drive exhaustion[SNS] and fatigue[SNS]. Those are

(1) the *overload* that arises from the ubiquitous nature of SNS, the interruptions that are its consequence, and the pace of change that they introduce;

(2) the *excessive use* of SNS;

(3) the *conflicts* that arise from the excessive use.

Table 7 summarizes the reported results so far and further illustrates the difference between the ICT use effects on the combined energy constructs (vigor$^{Mood}$, vigor$^{Work}$, and vitality$^{Mot}$) as well as the combined fatigue constructs (fatigue$^{Mood}$, exhaustion$^{Work}$, exhaustion$^{SNS}$, and fatigueSNS).

Based on the reported results, we come to six overarching conclusions. First, when researchers focus on the energy dimension,

they are slightly more likely to find that ICT use increases energy than that it depletes it (Table 7). The results are entirely different when the focus is on the fatigue dimension. Here, researchers are much more likely to find that ICT use increases fatigue than that it decreases it. The overall ratio between beneficial and detrimental effects on the energy dimension signaled a slightly beneficial relationship between ICT and energy (a bit higher than one, Table 7). On the other hand, the overall ratio of reported beneficial effects to detrimental effects on the fatigue dimension was almost 1:3 and strongly signaled a detrimental relationship (increased fatigue).

**Table 7. Summary of reported results that revealed a beneficial, neutral or detrimental effect of ICT use across constructs and aggregated for all energy and fatigue constructs**

| Energy constructs | | | | |
|---|---|---|---|---|
| Examined construct (Number of studies) | ICT use is beneficial | ICT use is neutral | ICT use is detrimental | Conditional/ Total effects |
| Vigor$^{Mood}$ (N=8) | 5 | 4 | 4 | 9/13 |
| Vigor$^{Work}$ (N = 10) | 15 | 14 | 12 | 40/41 |
| Vitality$^{Mot}$ (N=5) | 6 | 7 | 6 | 19/19 |
| **Total energy (N=23)** | **26** | **25** | **22** | **68/73** |
| Fatigue constructs | | | | |
| | ICT use is beneficial | ICT use is neutral | ICT use is detrimental | Conditional/ Total effects |
| Fatigue$^{Mood}$ (N=6) | 2 | 3 | 2 | 3/7 |
| Exhaustion$^{Work}$ (N=21) | 19 | 22 | 33 | 68/74 |
| Exhaustion$^{SNS}$ (N=8) | 1 | 3 | 20 | 23/24 |
| Fatigue$^{SNS}$ (N=6) | 1 | 2 | 12 | 15/15 |
| **Total fatigue (N=41)** | **23** | **30** | **67** | **109/120** |

The only exception of this negative ratio was fatigue$^{Mood}$, mostly due to wellness ICTs (Table 2). We conclude that if one follows the univariate paradigm, then the choice of construct—particularly whether it is an energy construct or a fatigue construct— plays a vital role in the obtained results. This discrepancy is evidently in line with the timely bivariate paradigm, but this paradigm was rarely followed as researchers were most likely to focus on only one dimension, and what is more, for it to be the fatigue dimension.

Second, we find that when researchers did follow the bivariate paradigm, i.e., looked at both dimensions simultaneously, they often found different results. The effect of ICT use on the fatigue dimension was stronger than on the energy dimension in five out of thirteen studies. This was true even in the rarer cases when the effect of ICT use on fatigue was beneficial (ICT use decreases fatigue). Moreover, even when the strength of the effects was similar in size (N=5), there were more reported effects for fatigue than for energy in the majority of those studies. This demonstrates that investigating the effects of ICT use solely on an energy dimension or a fatigue dimension does indeed give an incomplete picture.

Third, we find that whether ICT use fatigues or energizes users often depends on the context or the individual using them, i.e., the vast majority of the reported effects were conditional. The purpose of use, personality, self-evaluation, coping strategies, values, expectations, preferences and many other attitudes were all found to moderate the relationship between ICT use and the energy or fatigue outcomes. When this was the case, both detrimental and beneficial effects were reported on a single dimension, determined by the context or the individual. ICT use is not necessarily energizing or fatiguing in itself.

Fourth, when such mixed results were found, the detrimental effects were stronger than the beneficial in 35% of the cases. For comparison, the beneficial effects were stronger than the detrimental in 8% of the cases. This means that when different effects were found, it was over four times more likely that the detrimental effects were stronger than the beneficial. Moreover, in the studies that found equal strengths for both effects, the number of detrimental effects was bigger than the number of beneficial effects in the majority of cases, and especially if the differences were found on the fatigue dimension. Thus, even though ICT use might not always be fatiguing or de-energizing in itself, there is a tendency for ICT use to fatigue and slightly de-energize.

Fifth, the results differ considerably based on the type of ICT used (Table 8). For example, positive effects were mostly evident for welness ICTs, i.e., technology that was specifically designed to increase energy or decrease fatigue. On the other hand, conventional ICTs and especially social networks and cell phones, are most likely to have a negative effect on both energy and fatigue.

Finally, there are repeated patterns in the ways in which ICT use influences energy and fatigue across all constructs. This allows us to conclude our summary with a conceptual model and move towards a theory of the mechanisms through which ICT use can alter the experiences of subjective energy and fatigue.

# CONCEPTUAL FRAMEWORK FOR THE EFFECTS OF ICT USE ON ENERGY AND FATIGUE

| Examined construct | Excessive use | Overload | Autonomy | Conflict |
|---|---|---|---|---|
| **Table 9. Number of studies that reported effects conditional on excessive use, autonomy, overload, conflict and individual characteristics per construct** | | | | |
| Vigor$^{Mood}$ | 2 | | | |
| Fatigue$^{Mood}$ | 1 | | 1 | |
| Exhaustion$^{Work}$ | 3 | 4 | 7 | 7 |
| Vigor$^{Work}$ | 2 | | 3 | 1 |
| Vitality$^{Mot}$ | 2 | | | |
| Exhaustion$^{SNS}$ | 4 | 5 | | 1 |
| Fatigue$^{SNS}$ | | 3 | | |
| **Total** | **14** | **12** | **11** | **9** |

Considering the recurring patterns in the studies, regardless of construct and beyond individual personality, we observe four overarching and repeated themes for fatigue and three for energy:

(1) excessive ICT use;
(2) overload and pressure due to ICT use (only for fatigue);
(3) autonomy (self-determined ICT use and perceived autonomy); and
(4) experience of conflict (Table 9).

Whereas excessive (or problematic) use is just a subtype of ICT use and can thus be interpreted as the independent variable in our research question, there is a qualitative difference between ICT use and excessive ICT use beyond a mere increased frequency of use. It is also important to highlight excessive ICT use because it was the most frequently found repeated condition for energy depletion and fatigue increase in the review (Table 9). Moreover, excessive or problematic ICT use was reported for all studied constructs except fatigue$^{SNS}$. When the threshold between ICT use and excessive ICT use is crossed, ICT use becomes problematic and has a detrimental effect on transient and more chronic experiences of both subjective energy and fatigue.

A big driver of excessive ICT use is ICT's ubiquity. Ubiquitous ICTs expose people to a state of permanent availability, increased interruptions and via changed expectations, it "forces" people to (over) use their ICTs. Having mobile SNS for example, paired with a large number of SNS friends, increases the chances of excessive use (Maier et al. 2015). Even without ubiquity, people use ICTs when they feel low on energy because they believe ICT use will increase their energy (Aghaz and Sheikh 2016; Myrick 2015; Quinones and Griffiths 2017), suggesting a reverse causation between increased ICT use and energy. Consequently, when people use ICTs to recover energy, they might inadvertently increase their ICT use, which can in turn lower their energy, forming a self-reinforcing negative cycle.

An important consequence of (excessive) ICT use and ICT ubiquity is overload. Overload was reported as an important predictor in a quarter of all studies, notably only for the fatigue constructs (Table 9). For example, both ubiquity and (excessive) ICT use were shown to create work, information or interruption overload, email and mobile messenger overload, social and communication overload, etc. Some ICT design features such as complexity, high pace of change, technological failures, lack of support etc., were also linked to overload. All these types of "techno-overload" were repeatedly related to exhaustion or fatigue and were sometimes even operationalized in energy terms. For instance, the item "During my time off, work-related interruptions take up more energy than I have" measures interruption overload (Chen and Karahanna 2018).

The third important pattern that emerges from the review is that of autonomy and self-determined use. Even though excessive, i.e., problematic use is implicitly a non-autonomous behavior in itself, autonomy and self-determined use were explicitly singled out as a conditional or indirect driver of both energy and fatigue. In an organizational context, when employees feel more autonomy over their ICT use or because of the ICT use (for example in the case of telework or non-mandatory organizational IS use), they report more work vigor and less fatigue and exhaustion. Furthermore, employees who prefer to use work-related ICTs outside of working hours, who can control their work-life boundary and when to respond to messages, or those whose values are in line with the organizational ICTs, are more likely to feel energized and less likely to feel exhausted by their work and their jobs. In all mentioned cases, it is the self-determined, voluntary ICT use that relates to higher energy and lower fatigue.

Furthermore, in some of the studies that reported on excessive use, it was explicitly the perceived control, i.e., control failure, that determined whether ICTs had beneficial or detrimental effects on energy and fatigue. Thus, autonomy is the binding condition for ICT use to energize and reduce the fatigue of the ICT users, and might even be the qualitative difference between high frequency of use and problematic ICT use.

The fourth important mechanism that re-emerged across studies and constructs was conflict. ICT use can create different types of conflict, mostly through increased overload and diminished autonomy, but also through excessive use itself. For example, ICT control failure is in itself a technology-personal conflict. Furthermore, complex ICTs, or highly interrupting ICT systems can create technology-work conflicts. Overload also creates technology-family conflicts or work-life conflicts, as well as role conflicts. All forms of conflict are stressful events, and like overload, have been repeatedly linked to increased fatigue and even lowered energy.

Finally, a reverse link between fatigue and ICT use might also exist. This was shown in the case of both employees who are exhausted by their jobs and SNS users who are consciously exhausted or fatigued by their SNS use. In the case of employees, this leads to less frequent organizational ICT use, whereas in the case of SNS users, it leads to an intention to discontinue SNS use.

We summarize our findings in the conceptual model presented in Figure 2. Through it, we suggest that ICT use can have a small direct relationship with both energy (vigor, vitality) and fatigue, but most of the influence on these two experiences is due to autonomy and conflict, and in the case of fatigue, overwhelmingly through overload. This shows that some of the mechanisms through which ICT use fatigues users are different from those that energize. Finally, the strength with which these mediators and moderators operate on the two dimensions also differs (bolder lines in Figure 2 signify stronger relationships). For example, the experience of conflict due to ICT use has a much stronger effect on fatigue than on energy.

The model can also explain what we coin as "Digital Fatigue Paradox." Such a paradox exists when ICT use invigorates and fatigues at the same time or in parallel. The paradox was sometimes observed in studies that measured both an energy and a fatigue construct and almost always found a different result on the separate dimensions (Hancock 2007; Rhee and Kim 2016; Sardeshmukh et al. 2012; van Zoonen and Rice 2017). For example, Rhee and Kim (2016) found that using smartphones on work breaks can increase both vigor$^{Mood}$ and vigor$^{Work}$, but at the same time it can increase fatigue$^{Mood}$ and exhaustion$^{Work}$. This might happen because cell phone use creates overload, which only affects fatigue$^{Mood}$ and exhaustion$^{Work}$ but not vigor$^{Mood}$ and vigor$^{Work}$.

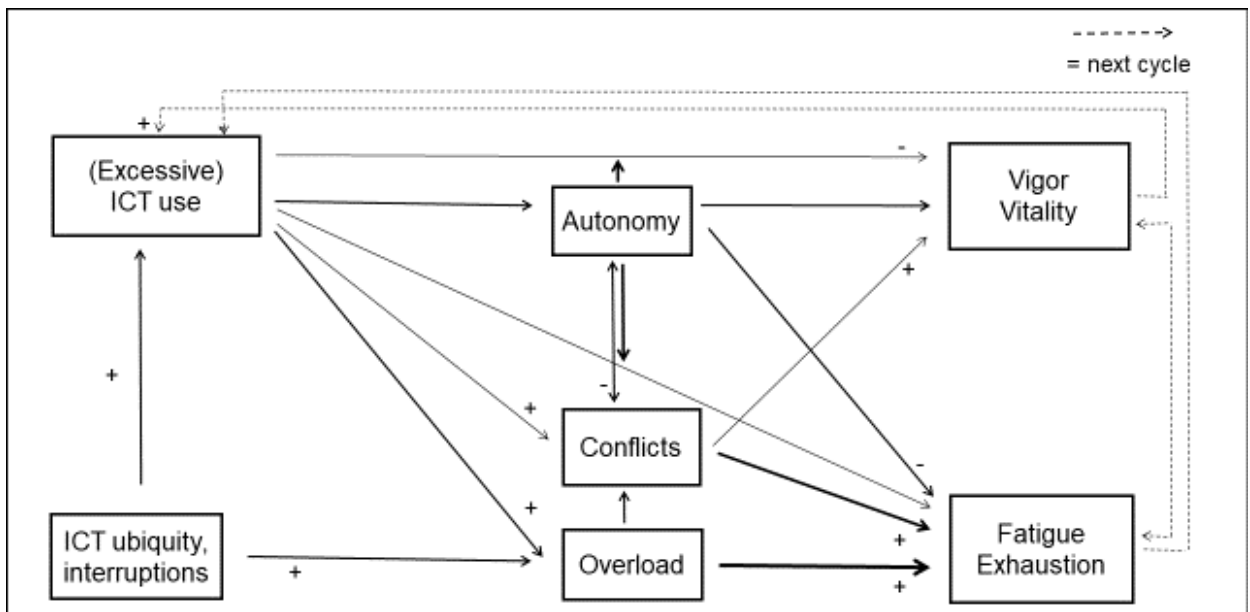

**Figure 2. Conceptual framework of the mechanisms through which ICT use energizes and fatigues users**

## DISCUSSION

The aim of the present study was to shed light on the current state of the art between ICT use and subjective energy and fatigue, i.e., to answer whether ICT use invigorates or fatigues its users. For this purpose, we first needed a conceptual clarity about the constructs subjective energy and subjective fatigue as well as regarding the paradigms often used in delineating their relationship. We provide such clarity by consolidating the diverse and sometimes ambiguous energy and fatigue constructs into four areas and seven construct categories based on the scientific field of origin (area) and operationalization (high-level construct, Table 1). We further discuss the implications of the bivariate and univariate paradigms of energy and fatigue, i.e., the consequences of treating the experiences of energy and fatigue as separate constructs, as opposed to two poles on one energy-fatigue construct. By following the timely bivariate paradigm, we contribute to the research by untangling the effects of ICT use on each dimension separately.

We then systematically review the literature on the relationship between ICT use and the seven different constructs, before summarizing the results in a conceptual framework (Figure 2). Whereas the review preliminary answers our research question, the conceptual model moves one step beyond and discusses the mechanisms through which such ICT effects on energy and fatigue are possible.

### Is ICT Use Energizing or Fatiguing

We find that the current state of the art on the relationship between ICT use and subjective energy or fatigue suggests a predominantly negative picture, especially on the fatigue dimension but also to a degree on the energy dimension (Table 7). Multiple studies have shown that ICT use causes chronic emotional exhaustion with work, but also depletes momentary feelings of energy, increases the mood of fatigue, and is negatively related to the more permanent states of work-related vigor and life vitality. In the case of social networking sites (SNS) and messengers, the experiences of exhaustion and fatigue were directly related to the use of the SNS themselves.

Nevertheless, our analysis shows that technology can have an energizing effect in all conceptualizations of subjective energy too, and it can decrease fatigue and exhaustion as well (Table 7). As a result, users often turn to ICTs when they are low on energy or when they simply want to take a break, relax or recover. When this happens, at least some users can successfully replenish energy through ICT use.

Thus, there appears to exist a "Digital Fatigue Paradox": the same technology that energizes users can cause fatigue too, sometimes potentially in parallel. We suggest that this is due to the different effects that ICT use has on the energy and fatigue dimensions, i.e., the different paths through which ICT use affects energy and fatigue (Figure 2).

The implication of the Digital Fatigue Paradox is that in a search for energy recovery, users might be leaving themselves vulnerable to fatiguing effects. Our review revealed more frequent detrimental effects on the fatigue dimension than on the energy dimension and additionally the effects on fatigue were stronger. When beneficial and detrimental effects were found on a single dimension, regardless of whether it is fatigue or energy, the strength of the detrimental effect was also stronger. This leads us to conclude that the current state of the art shows ICT use to have stronger fatiguing than invigorating effects on its users, even though the users might actually believe the opposite to be true.

### The Mechanisms through Which ICT use Energizes and Fatigues

Our conceptual model proposes that the invigorating effect mainly emerges when ICTs foster autonomy and are used voluntarily (autonomously), but also because ICT use satisfies other human needs, including the need for relaxation (direct path, Figure 2). This is why people voluntarily turn to ICTs when they feel low on energy and often experience increased vigor due to ICT use.

Teleworking and user system control are two examples where ICTs enable increased autonomy. Conversely, energy is depleted when users feel in any way controlled or forced (non-autonomous) to use ICTs, such as when they are expected to immediately respond to messages or when they are forced to use organizational ICTs that are in conflict with their values. Similarly, when the frequency of ICT use exceeds the subjective norm of the individual, a personal conflict arises that further decreases energy. These findings are in agreement with self-determination theory, which postulates that satisfying the need for autonomy enables eudemonic well-being characterized by the experience of energy (Ryan et al. 2008), and in organizational context, with the Job Demands-Resources Model, which sees autonomy as the core buffer against exhaustion[Work] (Bakker and Demerouti 2007).

On the other hand, ICT use fatigues its users predominantly through overload and conflict. Even when there is no perceivable ICT use problem and the use is autonomous —in which case ICT use increases vigor—the sheer ubiquity of ICTs, and the interruptions that come with it, have a potential to create different types of overload and consequently different types of conflict. Both overload and conflict have repeatedly proven to cause ICT exhaustion and fatigue. This is the core of the Digital Fatigue Paradox. While voluntary ICT use can energize, it can also cause fatigue in parallel. This is again in line with the Job Demands-Resources Model for organizational contexts, which postulates that overload and conflict are the main work stressors that cause exhaustion[Work] (Bakker and Demerouti 2007).

A reverse causation is also possible. Because of the energizing effect, users might increasingly use those particular ICTs that invigorate. This might prompt and habituate excessive use, since vigor is a pleasant affect.

It is no wonder that people turn to ICTs when they feel low on energy in many different situations: during work hours (cyberloafing), on work-breaks, in schools, as well as in their free time. The excessive use of ICTs, however, can start a fatiguing cycle through overload, including accumulating workload, and conflict, such as guilt, family conflict, or control failure. Even if one can control the (pleasant) ICT use, consciously refraining from going online or from responding immediately to messages can strain the cognitive resources, cause fatigue and in the long run turn into exhaustion.

A reverse causation is possible on the fatigue dimension as well. Exhausted employees avoid using organizational ICTs, whereas SNS users with increased SNS exhaustion tend to discontinue using those SNS (Bala and Bhagwatwar 2018; Cao and Sun 2018; Gao et al. 2018; Lo 2019; Luqman et al. 2017; Maier et al. 2015; Shin and Shin 2016; Zhang et al. 2016). This is in line with conservation of resources theory, which posits that initial loss of resources will prompt defensive attempts to conserve and protect the remaining ones (Hobfoll 1989). These findings highlight the possibility for grave consequences for both organizations and ICT platforms, i.e., they highlight the importance of finding the right balance between ICT use and human energy and fatigue through both policy and design.

## Implications and Organizational Policy Recommendations

Our review and conceptual framework allow for a number of design and policy recommendations for organizations. First, any design or organizational policy that hinders users' autonomy in relation to ICT use is likely to decrease energy and increase fatigue of employees, leading to devastating consequences for the organization. As a consequence, fatigued employees are likely to avoid using organizational ICTs and cyberloaf, exacerbating the negative effects of diminished autonomy even further.

Technology increases user autonomy when it is designed with absolute user control, but also more generally, when it allows for task control. On the other hand, organizational policy fosters autonomy when it gives the employees the freedom to choose where, when and how they use the organizational ICTs, especially after work hours.

Thus, mandatory introduction of technology, especially when coupled with inadequate training, contribute to employees' emotional exhaustion. Such introductions should invest time in explaining the value of the new ICT, i.e., training should not only focus on how-to but also on fostering voluntary use and aligning employees' values with the use of the new technology.

Our review also suggests that if possible, organizations should strongly consider allowing telework, at least part-time. Telework increases autonomy and task-control and is shown to both increase work vigor and decrease exhaustion. We are witnessing in the light of the Covid-19 pandemic that many employers are not comfortable with this option and have examined options to surveil their employees while they work from home. Recent reports show that remote staff is working almost an hour longer than usual during the pandemic because expectations that employees always be online have increased (Friedman 2020; Yeung 2021).

Our review suggests that this is a colossal mistake, since surveillance directly robs employees of their autonomy and task control and will thus only reverse the positive effects of telework, increase employees' exhaustion and deprive them of energy and work engagement. The first signs that this is happening are already there (Hern 2020).

Organizations should further strongly discourage work-related ICT use outside of working hours through building an organizational culture that is respectful of the employees' private time, especially in extraordinary times such as pandemics, where the work-life boundaries have been blurred even further. Still, after-hour work-related ICT use should be allowed, i.e., it shouldn't be outright prohibited or technically restricted. In this way, the energy draining, i.e., fatiguing effects of after-hour work-related ICT use will be avoided, while not robbing the employees of their autonomy in voluntarily engaging in after-hour work if they so wish.

Finally, organizations should be aware that ICT use can cause overload, which is a further strong driver of exhaustion. Overusing videoconferencing tools—which have already been reported to create so-called "Zoom fatigue" (Bailenson 2021)—should be avoided, and measures to decrease techno-overload should be seriously considered.

### Implications for Design of Conventional ICTs and Policy Recommendation for the Regulative Bodies

According to the SDT, any feeling of being pressured, coerced or even seduced by desired outcomes is a non-autonomous behavior (Deci and Ryan 2000). Consequently, seductive technology that is designed to "hook" and that is based on a business model dependent on excessive use (Eyal 2014; Zuboff 2019) fatigues its users more than technology with no such end goal.

Unfortunately, addictive technologies are more pervasive than ever, so perhaps unsurprisingly there is an increasing number of users, especially adolescents, who fall into the excessive use category. Studies suggest that depending on country, up to 38% of adolescents have a problematic relationship with the Internet, social media or their mobile phones (Durkee et al. 2012; Pedrero-Pérez et al. 2012; Vigna-Tagliantti et al. 2017). Thus, identifying non-functional, seductive design features and finding ways to regulate their use seems worth pursuing.

Self-regulatory efforts that aim at "removing pressure" to collect likes from SNS users, such as the one introduced by Instagram in Canada and Australia, are also welcomed, because they are aimed at reducing problematic use (Australian Associated Press 2019). The effects of such targeted changes (only a few countries are piloted at the moment) should be closely monitored, and if effective, they should give regulators the impetus to impose similar design changes to the industry as a whole, especially if they answer the questions of why and how regular ICT use turns into excessive use.

Finally, an attempt at categorization of the different technologies that may even give a human-centered seal of approval to the ones with a strong positive relationship with vitality and eudemonic well-being on the one hand, and a negative relationship with fatigue and exhaustion on the other hand, is well worth pursuing.

### Gaps in research and future outlook

As it stands, the current literature points to gaps in research that need to be closed in order to increase the confidence in our findings. First, the majority of studies were correlational, and only two tested for a reversed causation, which leaves open the important question of causality in the relationship between ICT use and human energy: Do people use certain technologies because they are tired or are they tired because they (over)use certain technologies? Or, is there a cycle of reversed causations? Different authors argued for different causation paths, which means that both are plausible, and in our conceptual model we suggest that a reverse causation is probable. Future research should try to clarify these questions and validate our model.

Second, the proposition that ICTs have different effects on the two human energy dimensions is heavily underresearched, as the univariate paradigm still predominates. The vast majority of the reviewed studies measured energy on only one dimension and many more not considered in this review either combine the energy and fatigue scales or measure arousal as a single dimension on pictorial or visual scales. Since energy and fatigue are shown to be bivariate, a lot of essential information is lost when researchers combine the scales or only look at one dimension, as we have shown throughout the review. Future studies should always use bivariate scales, especially when the variance in energy or fatigue is to be predicted by technology use. This opens up new questions such as can social media vigor be considered alongside social media fatigue. Studies of user experience on Facebook that used physiological correlates of arousal have shown that while on Facebook, users do exhibit an optimal physiological experience akin to *flow* (Cipresso et al. 2015; Mauri et al. 2011). If SNS vigor does exist and is researched together with SNS exhaustion, it might unravel more complicated dynamics of SNS use, beyond fatiguing effects.

Finally, our analysis shows that not all technologies have the same effects. Even though very few studies looked at the differences between different technologies directly, we can conclude that there is a general trend that links social media and (smart)phone use to fatigue and emotional exhaustion both in the private as well as in the organizational context (Gaudioso et al. 2017; Hietajärvi et al. 2019; van Zoonen and Rice 2017). However, since the majority of studies only referred to the Internet or ICTs in general, future research would greatly profit from distinguishing between different types of technology, different types of devices, different types of media as well as different types of media content.

*Conclusion*

ICT use can both energize and fatigue users, sometimes even at the same time, a phenomenon that we term Digital Fatigue Paradox. The main path through which ICTs invigorate is autonomy; the main paths through which they fatigue are overload and conflict. The current state of the art has more evidence for the fatiguing effect and it seems that this effect is stronger (Hancock 2007; Rhee and Kim 2016; Sardeshmukh et al. 2012; van Zoonen and Rice 2017). This means that even if ICT use is invigorating, the chances are high that their use can (also) cause fatigue. However, the focus on the energy dimension, and especially on both dimensions in parallel, was scarce, so more research is needed to confirm the invigorating effects of ICTs use.

**REFERENCES**

Aboujaoude, E. 2010. "Problematic Internet Use: An Overview," *World Psychiatry* (9:2), pp. 85-90.

Ackerman, P. L. 2011. *Cognitive Fatigue: Multidisciplinary Perspectives on Current Research and Future Applications*. American Psychological Association.

Aghaz, A., and Sheikh, A. 2016. "Cyberloafing and Job Burnout: An Investigation in the Knowledge-Intensive Sector," *Computers in Human Behavior* (62), pp. 51-60.

Agogo, D., and Hess, T. J. 2018. ""How Does Tech Make You Feel?" a Review and Examination of Negative Affective Responses to Technology Use," *European Journal of Information Systems* (27:5), pp. 570-599.

Akın, A. 2012. "The Relationships between Internet Addiction, Subjective Vitality, and Subjective Happiness," *CyberPsychology, Behavior & Social Networking* (15:8), pp. 404-410.

Aumayr-Pintar, C., Cerf, C., and Parent-Thirion, A. 2018. "Burnout in the Workplace: A Review of the Data and Policy Responses in the Eu,").

Australian Associated Press. 2019. "Instagram Hides Number of 'Likes' from Users in Australian Trial." from https://www.theguardian.com/technology/2019/jul/18/instagram-hides-number-of-likes-from-users-in-australian-trial

Ayyagari, R., Grover, V., and Purvis, R. 2011. "Technostress: Technological Antecedents and Implications," *MIS quarterly* (35:4), pp. 831-858.

Bailenson, J. N. 2021. "Nonverbal Overload: A Theoretical Argument for the Causes of Zoom Fatigue," *Technology, Mind, and Behavior* (2:1).

Bakker, A. B., and Demerouti, E. 2007. "The Job Demands-Resources Model: State of the Art," *Journal of managerial psychology* (22:3), pp. 309-328.

Bakker, A. B., Schaufeli, W. B., Leiter, M. P., and Taris, T. W. 2008. "Work Engagement: An Emerging Concept in Occupational Health Psychology," *Work & Stress* (22:3), pp. 187-200.

Bala, H., and Bhagwatwar, A. 2018. "Employee Dispositions to Job and Organization as Antecedents and Consequences of Information Systems Use," *Information Systems Journal* (28:4), pp. 650-683.

Baumeister, R. F. 2002. "Ego Depletion and Self-Control Failure: An Energy Model of the Self's Executive Function," *Self and identity* (1:2), pp. 129-136.

Bener, A., Yildirim, E., Torun, P., Çatan, F., Bolat, E., Alıç, S., Akyel, S., and Griffiths, M. D. 2018. "Internet Addiction, Fatigue, and Sleep Problems among Adolescent Students: A Large-Scale Study," *International Journal of Mental Health and Addiction*), pp. 1-11.

Berman, M. G., Jonides, J., and Kaplan, S. 2008. "The Cognitive Benefits of Interacting with Nature," *Psychological science* (19:12), pp. 1207-1212.

Berrios, G. E. 1990. "Feelings of Fatigue and Psychopathology: A Conceptual History," *Comprehensive psychiatry* (31:2), pp. 140-151.

Berto, R. 2005. "Exposure to Restorative Environments Helps Restore Attentional Capacity," *Journal of environmental psychology* (25:3), pp. 249-259.

Bialowolski, P., McNeely, E., VanderWeele, T. J., and Weziak-Bialowolska, D. 2020. "Ill Health and Distraction at Work: Costs and Drivers for Productivity Loss," *Plos one* (15:3), p. e0230562.

Boksem, M. A., and Tops, M. 2008. "Mental Fatigue: Costs and Benefits," *Brain research reviews* (59:1), pp. 125-139.

Bradley, M. M., and Lang, P. J. 1994. "Measuring Emotion: The Self-Assessment Manikin and the Semantic Differential," *Journal of behavior therapy and experimental psychiatry* (25:1), pp. 49-59.

Bright, L. F., Kleiser, S. B., and Grau, S. L. 2015. "Too Much Facebook? An Exploratory Examination of Social Media Fatigue," *Computers in Human Behavior* (44), pp. 148-155.

Bültmann, U., Kant, I., Kasl, S. V., Beurskens, A. J., and van den Brandt, P. A. 2002. "Fatigue and Psychological Distress in the Working Population: Psychometrics, Prevalence, and Correlates," *Journal of psychosomatic research* (52:6), pp. 445-452.

Cao, X., Masood, A., Luqman, A., and Ali, A. 2018. "Excessive Use of Mobile Social Networking Sites and Poor Academic Performance: Antecedents and Consequences from Stressor-Strain-Outcome Perspective," *Computers in Human Behavior* (85), pp. 163-174.

Cao, X., and Sun, J. 2018. "Exploring the Effect of Overload on the Discontinuous Intention of Social Media Users: An Sor Perspective," *Computers in human behavior* (81), pp. 10-18.

Chen, A., and Karahanna, E. 2018. "Life Interrupted: The Effects of Technology-Mediated Work Interruptions on Work and Nonwork Outcomes," *MIS Quarterly* (42:4), pp. 1023-1042.

Chen, I.-S. 2017. "Work Engagement and Its Antecedents and Consequences: A Case of Lecturers Teaching Synchronous Distance Education Courses," *Computers in Human Behavior* (72), pp. 655-663.

Cipresso, P., Serino, S., Gaggioli, A., Albani, G., Mauro, A., and Riva, G. 2015. "Psychometric Modeling of the Pervasive Use of Facebook through Psychophysiological Measures: Stress or Optimal Experience?," *Computers in Human Behavior* (49), pp. 576-587.

Cohen, J. 1988. *Statistical Power Analysis for the Behavioral Sciences*. Hillsdale, NJ, USA: Lawrence Erlbaum Associates.

Collins, R. 1990. "Stratification, Emotional Energy, and the Transient Emotions," *Research agendas in the sociology of emotions*), pp. 27-57.

Csikszentmihalyi, M. 1991. *Flow: The Psychology of Optimal Experience*. New York: Harper Collins.

Deci, E. L., and Ryan, R. M. 2000. "The" What" and" Why" of Goal Pursuits: Human Needs and the Self-Determination of Behavior," *Psychological inquiry* (11:4), pp. 227-268.

Deci, E. L., and Ryan, R. M. 2011. "Self-Determination Theory," *Handbook of theories of social psychology* (1:2011), pp. 416-433.

DeLuca, J., Genova, H. M., Capili, E. J., and Wylie, G. R. 2009. "Functional Neuroimaging of Fatigue," *Physical medicine and rehabilitation clinics of North America* (20:2), pp. 325-337.

Demerouti, E., and Bakker, A. B. 2008. "The Oldenburg Burnout Inventory: A Good Alternative to Measure Burnout and Engagement," *Handbook of stress and burnout in health care*), pp. 65-78.

Du, J., van Koningsbruggen, G. M., and Kerkhof, P. 2018. "A Brief Measure of Social Media Self-Control Failure," *Computers in Human Behavior* (84), pp. 68-75.

Durkee, T., Kaess, M., Carli, V., Parzer, P., Wasserman, C., Floderus, B., Apter, A., Balazs, J., Barzilay, S., and Bobes, J. 2012. "Prevalence of Pathological Internet Use among Adolescents in E Urope: Demographic and Social Factors," *Addiction* (107:12), pp. 2210-2222.

Eyal, N. 2014. *Hooked: How to Build Habit-Forming Products*. Penguin UK.

Faul, F., Erdfelder, E., Buchner, A., & Lang, A.-G. 2009. "Statistical Power Analyses Using G*Power 3.1: Tests for Correlation and Regression Analyses," *Behavior Research Methods* (41), pp. 1149-1160.

Friedman, A. 2020. "Proof Our Work-Life Balance Is in Danger (but There's Still Hope)." Retrieved 8,4.2021, 2021, from https://www.atlassian.com/blog/teamwork/data-analysis-length-of-workday-covid

Gao, W., Liu, Z., Guo, Q., and Li, X. 2018. "The Dark Side of Ubiquitous Connectivity in Smartphone-Based Sns: An Integrated Model from Information Perspective," *Computers in Human Behavior* (84), pp. 185-193.

Gaudioso, F., Turel, O., and Galimberti, C. 2017. "The Mediating Roles of Strain Facets and Coping Strategies in Translating Techno-Stressors into Adverse Job Outcomes," *Computers in Human Behavior* (69), pp. 189-196.

Guerrero-Bote, V. P., and Moya-Anegón, F. 2012. "A Further Step Forward in Measuring Journals' Scientific Prestige: The Sjr2 Indicator," *Journal of informetrics* (6:4), pp. 674-688.

Han, S., Shanafelt, T. D., Sinsky, C. A., Awad, K. M., Dyrbye, L. N., Fiscus, L. C., Trockel, M., and Goh, J. 2019. "Estimating the Attributable Cost of Physician Burnout in the United States," *Annals of internal medicine* (170:11), pp. 784-790.

Hancock, P. A. 2007. "On the Process of Automation Transition in Multitask Human–Machine Systems," *IEEE Transactions on Systems, Man, and Cybernetics-Part A: Systems and Humans* (37:4), pp. 586-598.

Hennington, A., Janz, B., and Poston, R. 2011. "I'm Just Burned Out: Understanding Information System Compatibility with Personal Values and Role-Based Stress in a Nursing Context," *Computers in Human Behavior* (27:3), pp. 1238-1248.

Hern, A. 2020. "Microsoft Productivity Score Feature Criticised as Workplace Surveillance." 8.4.2021, from https://www.theguardian.com/technology/2020/nov/26/microsoft-productivity-score-feature-criticised-workplace-surveillance?CMP=Share_iOSApp_Other#top

Herrero, R., Garcia-Palacios, A., Castilla, D., Molinari, G., and Botella, C. 2014. "Virtual Reality for the Induction of Positive Emotions in the Treatment of Fibromyalgia: A Pilot Study over Acceptability, Satisfaction, and the Effect of Virtual Reality on Mood," *Cyberpsychology, Behavior, and Social Networking* (17:6), pp. 379-384.

Herzog, T. R., Maguire, P., and Nebel, M. B. 2003. "Assessing the Restorative Components of Environments," *Journal of Environmental Psychology* (23:2), pp. 159-170.

Hietajärvi, L., Salmela-Aro, K., Tuominen, H., Hakkarainen, K., and Lonka, K. 2019. "Beyond Screen Time: Multidimensionality of Socio-Digital Participation and Relations to Academic Well-Being in Three Educational Phases," *Computers in Human Behavior* (93), pp. 13-24.

Hobfoll, S. E. 1989. "Conservation of Resources: A New Attempt at Conceptualizing Stress," *American psychologist* (44:3), p. 513.

Huang, H.-C., Wong, M.-K., Yang, Y.-H., Chiu, H.-Y., and Teng, C.-I. 2017. "Impact of Playing Exergames on Mood States: A Randomized Controlled Trial," *CyberPsychology, Behavior & Social Networking* (20:4), pp. 246-250.

Jang, W. E., Bucy, E. P., and Cho, J. 2018. "Self-Esteem Moderates the Influence of Self-Presentation Style on Facebook Users' Sense of Subjective Well-Being," *Computers in Human Behavior* (85), pp. 190-199.

Jones, B. E. 2003. "Arousal Systems," *Front Biosci* (8:5), pp. 438-451.

Kelley, K. J., and Gruber, E. M. 2010. "Psychometric Properties of the Problematic Internet Use Questionnaire," *Computers in Human Behavior* (26:6), pp. 1838-1845.

Kuppens, P., Tuerlinckx, F., Russell, J. A., and Barrett, L. F. 2013. "The Relation between Valence and Arousal in Subjective Experience," *Psychological Bulletin* (139:4), p. 917.

Lee, A. R., Son, S.-M., and Kim, K. K. 2016. "Information and Communication Technology Overload and Social Networking Service Fatigue: A Stress Perspective," *Computers in Human Behavior* (55), pp. 51-61.

Lee, J. E., Xiang, P., and Gao, Z. 2017. "Acute Effect of Active Video Games on Older Children's Mood Change," *Computers in Human Behavior* (70), pp. 97-103.

Levy, Y., and Ellis, T. J. 2006. "A Systems Approach to Conduct an Effective Literature Review in Support of Information Systems Research," *Informing Science* (9).

Lieberman, H. R. 2007. "Cognitive Methods for Assessing Mental Energy," *Nutritional Neuroscience* (10:5-6), pp. 229-242.

Lim, M., and Yang, Y. 2015. "Effects of Users' Envy and Shame on Social Comparison That Occurs on Social Network Services," *Computers in Human Behavior* (51), pp. 300-311.

Lin, S. C., Tsai, K. W., Chen, M. W., and Koo, M. 2013. "Association between Fatigue and Internet Addiction in Female Hospital Nurses," *Journal of advanced nursing* (69:2), pp. 374-383.

Llorens, S., Schaufeli, W., Bakker, A., and Salanova, M. 2007. "Does a Positive Gain Spiral of Resources, Efficacy Beliefs and Engagement Exist?," *Computers in human behavior* (23:1), pp. 825-841.

Lo, J. 2019. "Exploring the Buffer Effect of Receiving Social Support on Lonely and Emotionally Unstable Social Networking Users," *Computers in Human Behavior* (90), pp. 103-116.

Luqman, A., Cao, X., Ali, A., Masood, A., and Yu, L. 2017. "Empirical Investigation of Facebook Discontinues Usage Intentions Based on Sor Paradigm," *Computers in Human Behavior* (70), pp. 544-555.

Magistretti, P. J., and Allaman, I. 2015. "A Cellular Perspective on Brain Energy Metabolism and Functional Imaging," *Neuron* (86:4), pp. 883-901.

Maier, C., Laumer, S., Eckhardt, A., and Weitzel, T. 2015. "Giving Too Much Social Support: Social Overload on Social Networking Sites," *European Journal of Information Systems* (24:5), pp. 447-464.

Mäkikangas, A., Kinnunen, S., Rantanen, J., Mauno, S., Tolvanen, A., and Bakker, A. B. 2014. "Association between Vigor and Exhaustion During the Workweek: A Person-Centered Approach to Daily Assessments," *Anxiety, Stress, & Coping* (27:5), pp. 555-575.

Marks, S. R. 1977. "Multiple Roles and Role Strain: Some Notes on Human Energy, Time and Commitment," *American sociological review*), pp. 921-936.

Maslach, C., Jackson, S. E., and Leiter, M. P. 1996. *Mbi: Maslach Burnout Inventory*. CPP, Incorporated Sunnyvale (CA).

Maslach, C., Jackson, S. E., Leiter, M. P., Schaufeli, W. B., and Schwab, R. L. 1986. *Maslach Burnout Inventory*. Consulting psychologists press Palo Alto, CA.

Maslach, C., Schaufeli, W. B., and Leiter, M. P. 2001. "Job Burnout," *Annual review of psychology* (52:1), pp. 397-422.

Mauri, M., Cipresso, P., Balgera, A., Villamira, M., and Riva, G. 2011. "Why Is Facebook So Successful? Psychophysiological Measures Describe a Core Flow State While Using Facebook," *Cyberpsychology, Behavior, and Social Networking* (14:12), pp. 723-731.

Mazmanian, M., Orlikowski, W. J., and Yates, J. 2013. "The Autonomy Paradox: The Implications of Mobile Email Devices for Knowledge Professionals," *Organization science* (24:5), pp. 1337-1357.

McArdle, W. D., Katch, F. I., and Katch, V. L. 1991. "Exercise Physiology: Energy, Nutrition, and Human Performance." LWW.

McNair, D. 1984. "This Week's Citation Classic," *Current contents* (27).

McNair, D., Lorr, M., and Droppleman, L. 1971. *Profile of Mood States (Poms)*. San Diego: Educational and Industrial Testing Service.

McNair, D., Lorr, M., and Droppleman, L. 1989. "Profile of Mood States (Poms),").

Moore, J. E. 2000. "One Road to Turnover: An Examination of Work Exhaustion in Technology Professionals," *MIS quarterly*), pp. 141-168.

Myrick, J. G. 2015. "Emotion Regulation, Procrastination, and Watching Cat Videos Online: Who Watches Internet Cats, Why, and to What Effect?," *Computers in human behavior* (52), pp. 168-176.

O'Connor, P. J. 2004. "Evaluation of Four Highly Cited Energy and Fatigue Mood Measures," *Journal of psychosomatic research* (57:5), pp. 435-441.

O'Connor, P. J. 2006. "Mental Energy: Assessing the Mood Dimension," *Nutrition reviews* (64:suppl_3), pp. S7-S9.

Ohly, S., and Latour, A. 2014. "Work-Related Smartphone Use and Well-Being in the Evening," *Journal of Personnel Psychology*).

Pedrero-Pérez, E., Rodríguez-Monje, M., and Ruiz Sánchez de León, J. 2012. "Mobile Phone Abuse or Addiction. A Review of the Literature," *Adicciones* (24), pp. 139-152.

Pendell, R. 2018. "Millennials Are Burning Out," Gallup, https://www.gallup.com/workplace/237377/millennials-burning.aspx.

Piszczek, M. M. 2017. "Boundary Control and Controlled Boundaries: Organizational Expectations for Technology Use at the Work–Family Interface," *Journal of Organizational Behavior* (38:4), pp. 592-611.

QSR International. 2018. "Nvivo." QSR International Pty Ltd, p. NVivo qualitative data analysis software.

Quinn, R. W., Spreitzer, G. M., and Lam, C. F. 2012. "Building a Sustainable Model of Human Energy in Organizations: Exploring the Critical Role of Resources," *The Academy of Management Annals* (6:1), pp. 337-396.

Quinones, C., and Griffiths, M. D. 2017. "The Impact of Daily Emotional Demands, Job Resources and Emotional Effort on Intensive Internet Use During and after Work," *Computers in Human Behavior* (76), pp. 561-575.

Ragsdale, J. M., and Hoover, C. S. 2016. "Cell Phones During Nonwork Time: A Source of Job Demands and Resources," *Computers in Human Behavior* (57), pp. 54-60.

Ragu-Nathan, T., Tarafdar, M., Ragu-Nathan, B. S., and Tu, Q. 2008. "The Consequences of Technostress for End Users in Organizations: Conceptual Development and Empirical Validation," *Information systems research* (19:4), pp. 417-433.

Rashid, T., and Asghar, H. M. 2016. "Technology Use, Self-Directed Learning, Student Engagement and Academic Performance: Examining the Interrelations," *Computers in Human Behavior* (63), pp. 604-612.

Reinecke, L., Hartmann, T., and Eden, A. 2014. "The Guilty Couch Potato: The Role of Ego Depletion in Reducing Recovery through Media Use," *Journal of Communication* (64:4), pp. 569-589.

Reinke, K., and Chamorro-Premuzic, T. 2014. "When Email Use Gets out of Control: Understanding the Relationship between Personality and Email Overload and Their Impact on Burnout and Work Engagement," *Computers in Human Behavior* (36), pp. 502-509.

Rhee, H., and Kim, S. 2016. "Effects of Breaks on Regaining Vitality at Work: An Empirical Comparison of 'Conventional' and 'Smart Phone' Breaks," *Computers in Human Behavior* (57), pp. 160-167.

Russell, J. A. 1980. "A Circumplex Model of Affect," *Journal of personality and social psychology* (39:6), p. 1161.

Ryan, R. M., and Deci, E. L. 2008. "From Ego Depletion to Vitality: Theory and Findings Concerning the Facilitation of Energy Available to the Self," *Social and Personality Psychology Compass* (2:2), pp. 702-717.

Ryan, R. M., and Frederick, C. 1997. "On Energy, Personality, and Health: Subjective Vitality as a Dynamic Reflection of Well-Being," *Journal of personality* (65:3), pp. 529-565.

Ryan, R. M., Huta, V., and Deci, E. L. 2008. "Living Well: A Self-Determination Theory Perspective on Eudaimonia," *Journal of happiness studies* (9:1), pp. 139-170.

Salanova, M., Grau, R. M., Cifre, E., and Llorens, S. 2000. "Computer Training, Frequency of Usage and Burnout: The Moderating Role of Computer Self-Efficacy," *Computers in Human Behavior* (16:6), pp. 575-590.

Sardeshmukh, S. R., Sharma, D., and Golden, T. D. 2012. "Impact of Telework on Exhaustion and Job Engagement: A Job Demands and Job Resources Model," *New Technology, Work and Employment* (27:3), pp. 193-207.

Satici, S. A., and Uysal, R. 2015. "Well-Being and Problematic Facebook Use," *Computers in Human Behavior* (49), pp. 185-190.

Schaufeli, W., Leiter, M., Maslach, C., and Jackson, S. 1996. "Mbi-General Survey (Mbi-Gs)," *Palo Alto, CA: Mindgarden*).

Schaufeli, W., Salanova, M., González-Romá, V., and Bakker, A. B. 2002. "The Measurement of Engagement and Burnout: A Two Sample Confirmatory Factor Analytic Approach," *Journal of Happiness Studies* (3:1), pp. 71-92.

Schaufeli, W. B., Bakker, A. B., and Salanova, M. 2006. "The Measurement of Work Engagement with a Short Questionnaire: A Cross-National Study," *Educational and psychological measurement* (66:4), pp. 701-716.

Shacham, S. 1983. "A Shortened Version of the Profile of Mood States," *Journal of personality assessment* (47:3), pp. 305-306.

Shanafelt, T. D., West, C. P., Sinsky, C., Trockel, M., Tutty, M., Satele, D. V., Carlasare, L. E., and Dyrbye, L. N. 2019. "Changes in Burnout and Satisfaction with Work-Life Integration in Physicians and the General Us Working Population between 2011 and 2017," *Mayo Clinic Proceedings*: Elsevier.

Shin, J., and Shin, M. 2016. "To Be Connected or Not to Be Connected? Mobile Messenger Overload, Fatigue, and Mobile Shunning," *CyberPsychology, Behavior & Social Networking* (19:10), pp. 579-586.

Shirom, A. 2011. "Vigor as a Positive Affect at Work: Conceptualizing Vigor, Its Relations with Related Constructs, and Its Antecedents and Consequences," *Review of General Psychology* (15:1), pp. 50-64.

Takeuchi, H., Taki, Y., Sekiguchi, A., Nouchi, R., Kotozaki, Y., Nakagawa, S., Miyauchi, C. M., Iizuka, K., Yokoyama, R., and Shinada, T. 2017. "Mean Diffusivity of Basal Ganglia and Thalamus Specifically Associated with Motivational States among Mood States," *Brain Structure and Function* (222:2), pp. 1027-1037.

Thayer, R. E. 1986. "Activation-Deactivation Adjective Check List: Current Overview and Structural Analysis," *Psychological reports* (58:2), pp. 607-614.

Thayer, R. E. 1990. *The Biopsychology of Mood and Arousal*. Oxford University Press.

van Zoonen, W., and Rice, R. E. 2017. "Paradoxical Implications of Personal Social Media Use for Work," *New Technology, Work and Employment* (32:3), pp. 228-246.

Vigna-Taglianti, F., Brambilla, R., Priotto, B., Angelino, R., Cuomo, G., and Diecidue, R. 2017. "Problematic Internet Use among High School Students: Prevalence, Associated Factors and Gender Differences," *Psychiatry research* (257), pp. 163-171.

Ware Jr, J. E., and Sherbourne, C. D. 1992. "The Mos 36-Item Short-Form Health Survey (Sf-36): I. Conceptual Framework and Item Selection," *Medical care*), pp. 473-483.

Windeler, J. B., Chudoba, K. M., and Sundrup, R. Z. 2017. "Getting Away from Them All: Managing Exhaustion from Social Interaction with Telework," *Journal of Organizational Behavior* (38:7), pp. 977-995.

Xie, J., Ma, H., Zhou, Z. E., and Tang, H. 2018. "Work-Related Use of Information and Communication Technologies after Hours (W_Icts) and Emotional Exhaustion: A Mediated Moderation Model," *Computers in Human Behavior* (79), pp. 94-104.

Yeung, P. 2021. "'If You Switch Off, People Think You're Lazy': Demands Grow for a Right to Disconnect from Work " Retrieved 8.4.2021, from https://www.theguardian.com/world/2021/feb/10/if-you-switch-off-people-think-youre-lazy-demands-grow-for-a-right-to-disconnect-from-work

Zhang, S., Zhao, L., Lu, Y., and Yang, J. 2016. "Do You Get Tired of Socializing? An Empirical Explanation of Discontinuous Usage Behaviour in Social Network Services," *Information & Management* (53:7), pp. 904-914.

Zheng, X., and Lee, M. K. 2016. "Excessive Use of Mobile Social Networking Sites: Negative Consequences on Individuals," *Computers in Human Behavior* (65), pp. 65-76.

Zuboff, S. 2019. *The Age of Surveillance Capitalism: The Fight for a Human Future at the New Frontier of Power*. Profile Books.

# APPENDIX

**Descriptive summary of the Reviewed Studies: Construct of Interest, Instrument, Method and Sample Used, Main Findings and Effect Size of Results**

| Table A1. Empirical studies on the relationship between technology use and subjective energy | | | | | | |
|---|---|---|---|---|---|---|
| **Study Number** | **Subjective Energy Term** | **Instrument** | **Methodology, Method** | **Sample** | **Main Findings** | **Effect size** |
| #1 Aghaz and Sheikh (2016) | Work-related emotional exhaustion (as part of job burnout) | MBI | Survey study Structural equation modeling | 298 employees in 5 knowledge-intensive firms | Job burnout is related to both cyberloafing activities and cyberloafing behaviors <br><br> The strongest reason for cyberloafing is recovery (to recover, r=.84) and because of addiction (r=.71), much less as a deviant organizational behavior (r=.52) or for learning (r=.50) The most common cyberloafing activities are social (r =.68) and informational (.r=.63) | ß =.28 ß =.47 |
| #2 Akın (2012) | Subjective vitality | SVS | Survey study Hierarchical regression | 328 university students | Internet addiction is negatively related to vitality (and it further mediates the link between Internet addiction and happiness) | ß = − .51 |
| #3 Ayyagari et al. (2011) | Work-related emotional exhaustion (strain) | Adopted MBI | Survey study Structural equation modeling | 661 employees | Presenteeism (reachability) indirectly increases exhaustion through work-home conflict, work overload and role ambiguity (all directly measured as consequences of ICT use) Pace of change indirectly increases exhaustion through work overload, role ambiguity and job insecurity | ß = .52, ß = .17 ß = .61, ß = .26 ß = .61, ß = .27 ß = .14, ß = .26 ß = .23, ß = .27 ß = .14, ß = .10 |

| Study Number | Subjective Energy Term | Instrument | Methodology, Method | Sample | Main Findings | Effect size |
|---|---|---|---|---|---|---|
| #4 Bala and Bhagwatwar (2018) #4 Bala and Bhagwatwar (2018) | Work-related emotional exhaustion | Adopted MBI | Study 1: Longitudinal study, 3-wave survey study; Pre-system implementation, 3 months and 6 months post-implementation. System = functional health and safety management system | 257 employees from a large manufacturing firm | Study 1. Exhaustion before system implementation and training (T1) negatively predicts the lean system use (frequency, duration and intensity of use) and the cognitive absorption use (i.e., when the employee uses the system without distractions and can immerse in the tasks), but not deep structure use (i.e., when the employee routinely uses the system features that were thought in training) 6 months later (T3) both<br><br>Deep structure use and cognitive absorption use at T2 are *negatively related* to exhaustion at T2. Lean structure use is not related. | ß = − .14<br>ß = − .16<br>ß = − .04<br><br><br><br>ß = − .12<br>ß = − .19<br>ß = − .06 |
| | Work-related emotional exhaustion | Adopted MBI | Study 2: Longitudinal study, 3-wave survey study; Pre-system implementation, 3 months and 6 months post-implementation System = enterprise system<br><br>Structural equation modeling | 181 employees from the same manufacturing firm | Study 2. Exhaustion before system implementation and training (T1) negatively predicts both the lean system use and the cognitive absorption use but not deep structure use 6 months later (T2)<br><br>Deep structure use at T2 is positively related to exhaustion, cognitive absorption use and lean structure use are not related | ß = − .15<br>ß = − .18<br>ß = − .11<br><br><br>ß = .28<br>ß = − .08<br>ß = .07 |

| Study Number | Subjective Energy Term | Instrument | Methodology, Method | Sample | Main Findings | Effect size |
|---|---|---|---|---|---|---|
| #5 Botella et al. (2016) | Vigor-activity (mood) Fatigue-inertia (mood) | POMS, monthly | Longitudinal study; 4 months long spaceship simulator study for astronauts with a mood-inducing VR. Alternated between mandatory VR use for a month (min 3 times a week) vs. a month of non-use. 2 monthly measures (before and after VR intervention); Did not report method of analysis; but did present raw data | 6 astronauts | VR use did not impact month-long energy VR use did not impact month-long fatigue | n.s. |
| #6 Bright et al. (2015) | SNS fatigue | Self-developed social media fatigue scale | Survey study Regression analysis | 747 SNS users with an active Facebook account | Social media privacy concerns, helpfulness, and SN self-efficacy are positively related to SNS fatigue Social media confidence is negatively SNS fatigue Note. the items for SNS fatigue resemble items for information overload | ß =.45, ß =.20, ß =.22 ß = − .32 |
| #7 Cao et al. (2018) | Mobile SNS exhaustion (techno-exhaustion) | Adopted MBI for mobile SNS exhaustion | Survey (paper and pencil) study Structural equation modeling | 505 mobile SNS app users (students) | Excessive SNS use and cognitive emotional preoccupation with SNS increase SNS exhaustion Cognitive-behavioral control (awareness and attempt to self-regulate use) moderates both relationships, i.e., it weakens them | ß =.34 ß =.09 ß =.− .30 ß = − .15 |
| #8 Cao and Sun (2018) | SNS exhaustion | Adopted MBI for social media exhaustion | Survey study Structural equation modeling | 258 SNS users (students) | Information overload and social overload increase SNS exhaustion | ß =.26 ß =.37 |

| Study Number | Subjective Energy Term | Instrument | Methodology, Method | Sample | Main Findings | Effect size |
|---|---|---|---|---|---|---|
| #9 Chen and Karahanna (2018) | Work-related and non-work related emotional exhaustion | MBI | Survey study<br><br>Structural equation modeling | 237 knowledge workers | Extent of after-hours work-related interruptions indirectly increases work exhaustion through psychological transitions and interruption overload<br><br>Extent of after-hours work-related interruptions indirectly increases non work (private life) exhaustion through interruption overload<br><br>Extent of after-hours work-related interruptions indirectly decreases work exhaustion through task closure | ß =.72; ß =.48 (indirect effect ß =.34)<br>ß =.74; ß = .33 (indirect effect ß =.24)<br>ß =.74; ß = .29 (indirect effect ß =.21)<br>ß =.51; ß = − .18 (indirect effect ß =-.09) |
| #10 Chen (2017) | Work-related vigor (as part of work engagement) | UWES | Longitudinal study: Multi wave survey study: Baseline + 10 daily diary measures on 10 consecutive teaching days;<br><br>Multilevel regression analysis | 40 lecturers who teach synchronous distance courses | Technical problems with equipment reduce daily vigor; support stuff increases it | Coeff.=-.19; coeffe =.26 (not clear if its standardized) c |
| #11 Gao, Liu, Guo, and Li (2018) | SNS exhaustion | Adopted MBI for social media exhaustion | Survey study<br><br>Structural equation modeling | 528 SNS users | Ubiquitous connectivity causes SNS exhaustion directly, and indirectly through information overload | ß =.18; ß =.32, ß =.26 |
| #12 Gaudioso, Turel, and Galimberti (2017) | Work-related emotional exhaustion | MBI | Survey study<br><br>Structural equation modeling | 242 employees in a large government organization | Techno-invasion increases exhaustion trough work-family conflict and through maladaptive coping strategies<br><br>Adaptive strategies weaken the relationship but work-family conflict has stronger relationship with maladaptive than adaptive strategies | ß =.66; ß = .40; ß = 1.11<br><br>ß = −.63; ß = .16 vs. ß = .40<br><br>ß =.51; ß = .50; ß = 1.11 |

| Study Number | Subjective Energy Term | Instrument | Methodology, Method | Sample | Main Findings | Effect size |
|---|---|---|---|---|---|---|
| #12 Gaudioso, Turel, and Galimberti (2017) | Work-related emotional exhaustion | MBI | Survey study  Structural equation modeling | 242 employees in a large government organization | Techno-overload increases exhaustion trough distress (adaptive coping strategies weaken the relationship)  Adaptive strategies weaken the relationship but distress has stronger relationship with maladaptive than adaptive strategies | ß = –.63; ß = .21 vs. ß = .50 |
| #13 Hancock (2007) | Fatigue (mood) Vigor(mood) | POMS | Experiment: Manipulated level of automation/pilot control in a cockpit simulator; 4 groups  ANOVA | 30 experienced pilots | The pilots experienced progressive increase in fatigue as the degree of system control increased  There was no effect on vigor | Large effect size, derived from sample size and critical F value ($F$[3, 28] = 3.033, $p$ = 0.0457), |
| #14 Hennington, Janz, and Poston (2011) | Work-related emotional exhaustion | MBI | Survey study  Structural equation modeling | 71 nurses from a large urban hospital who have been using the IS (mandatory electronic medical record system) for 3 years | Incompatibility (ß reversed score for compatibility) of IS with personal values increases the emotional exhaustion through creating role conflict  Note that more nurses felt that ICT use was compatible with their values and thus had decreased exhaustion. | ß = –.29; ß = .57 |
| #15 Herrero et al. (2014) | Vigor/energy (mood)  Fatigue (mood) | 1 item for intensity of the emotion vigor; 1 item baseline fatigue 1 item post induction for change of fatigue | Experiment: Pre-test Post-test treatment survey with 2h long mood inducing VR;  t-tests | 40 female patients with fibromyalgia | VR MIP increased vigor, had no effect on reported fatigue, but the vast majority of patients still felt that the VR MIP made fatigue "somewhat better"(42%.), "better (11%) or much better (11%) | Cohen's d =.38 Cohen's d =.07 n.s. Cohen's d =6.25 |

| Study Number | Subjective Energy Term | Instrument | Methodology, Method | Sample | Main Findings | Effect size |
|---|---|---|---|---|---|---|
| #16 Hietajärvi, Salmela-Aro, Tuominen, Hakkarainen, and Lonka (2019) | Study exhaustion<br><br>Study vigor (as part of study engagement | Adopted MBI for students<br><br>Adopted UWES for students | Survey study<br><br>Structural equation modeling | 741 elementary school students from 33 schools; | Elementary school students: SNS oriented Internet use and social gaming increase exhaustion<br><br>Elementary school students: SNS oriented Internet use decreases engagement; knowledge oriented increases engagement | ß = .17, ß = .09<br><br>ß = −.22, ß = .22 |
| | Study exhaustion<br><br>Study vigor (as part of study engagement | Adopted MBI for students<br><br>Adopted UWES for students | Survey study<br><br>Structural equation modeling | 1371 high school students from 18 high schools | High school students: SNS, blogging and media oriented Internet use increase exhaustion; action gaming decreases it<br><br>High school students: Knowledge oriented Internet use and social gaming increase engagement; action gaming decreases it. | ß = .10, ß = .13, ß = .11; ß = −.12<br><br>ß = .18, ß = .09; ß = −.18 |
| | Study exhaustion<br><br>Study vigor (as part of study engagement | Adopted MBI for students<br><br>Adopted UWES for students | Survey study<br><br>Structural equation modeling | 1232 higher education students from 3 institutions | University students: SNS oriented Internet use as well as action and social gaming decrease engagement, knowledge oriented use increases it | ß = −.15, ß = −.25, ß = −.09, ß = .21 |
| #17 Huang et al. (2017) | Vigor (mood) | POMS | Experiment: Pre and post-intervention survey; control vs. intervention group; intervention = play randomly selected exergames for 30 consecutive minutes once a week for 2 weeks;<br><br>Repeated measures (RM) ANOVA | 168 intervention and 167 control group participants (university staff and students) | Playing exergames increased vigor (in comparison to control group) | NA |

| Study Number | Subjective Energy Term | Instrument | Methodology, Method | Sample | Main Findings | Effect size |
|---|---|---|---|---|---|---|
| #18 Ishii and Markman (2016) | Work-related emotional exhaustion | MBI | Survey study<br><br>Bivariate regression analysis | 130 IT online customer help desk employee (use phone, e-mail or chat to provide services9 | Remote help providers feel more emotional presence with customers when they use phone than when they use e-mail or chat. Emotional presence in turn decreases exhaustion. | r=-18<br>ß=-.25 |
| #19 James et al. (2019) | Subjective vitality | SVS | Survey study<br><br>Structural equation modeling | 880 fitness technology users (app and/or device) | Use of social interaction features of fitness tech moderates (increases) the relationship between intrinsic and integrated regulation and vitality, as well as non-regulation, but it decreases it Use of data management features of fitness tech moderates (increases) the relationship between intrinsic and integrated regulation and vitality, as well as non-regulation, but it decreases it for external and introjected regulation | NA<br><br>No coefficients reported, only confidence intervals of interaction effects. |
| #20 Jang et al. (2018) | Subjective vitality | SVS | Experiment 1: Participants were asked to update their FB profile by either writing about their true or their strategic selves. Post-experiment survey<br><br>Experiment 2: Participants were asked to share a life event on FB profile that either reflected their true or their strategic selves; Post-experiment survey<br><br>Regression analysis | 136 SNS users<br><br><br><br>146 SNS users | In both experiments strategic self-presentation (as opposed to authentic) had no effect on subjective vitality | n.s.<br><br><br><br>n.s. |

| Study Number | Subjective Energy Term | Instrument | Methodology, Method | Sample | Main Findings | Effect size |
|---|---|---|---|---|---|---|
| #21<br>A. R. Lee et al. (2016) | SNS fatigue | Self-developed SNS fatigue scale | Survey (online and offline) study<br><br>Structural equation modeling | 250 university students | Information equivocality increases SNS fatigue through information overload<br><br>System pace of change increases SNS fatigue through system feature overload<br><br>System complexity increases SNS fatigue through system feature overload<br><br>Communication overload increases SNS fatigue | ß =.12, ß =.25<br><br>ß =.12, ß =.25<br><br>ß =.57, ß =.25<br><br>ß =.23 |
| #22<br>J. E. Lee et al. (2017) | Vigor (mood)<br><br>Fatigue (mood) | Adopted POMS | Experiment: Pre-Post intervention survey. Intervention: 30 min Active video game (AVG) session<br><br>ANOVA, MANOVA | 134 elementary school children (8-11 years old) | AVG session decreased the vigor and the fatigue of the school students, but the results for fatigue did not reach significance | $\eta_p^2$ = −0.38<br>$\eta_p^2$ = 0.06 |
| #23<br>Lim and Yang (2015) | SNS exhaustion (burnout) | Adopted MBI for social media exhaustion | Survey study<br><br>Structural equation modeling | 446 SNS users | Social comparison increases SNS exhaustion directly, and through increase of shame | ß =.33<br>ß =.43, ß =.50 |
| #24<br>Llorens et al. (2007) | (student) Vigor | UWES | Longitudinal study: 2-wave study (3-weeks apart) where participants were solving group task exclusively via chat (mIRC), surveys pre and post task solving<br><br>Structural equation modeling | 110 psychology students | Task resources (time and method control = perceived autonomy inherit in chat technology) increase efficacy beliefs which in turn increase vigor<br><br>Reverse causation is also present! (vigor at T1 to efficacy-beliefs at T2, efficacy beliefs at T2 to vigor at T2) | Cross-sect:<br>T1: ß =.44, ß =.61<br>T2: ß =.28, ß =.23<br>Long:<br>T1->T2:<br>ß =.20, ß =.50<br><br>T1<br>ß =30, ß =.23 |

| Study Number | Subjective Energy Term | Instrument | Methodology, Method | Sample | Main Findings | Effect size |
|---|---|---|---|---|---|---|
| #25 Lo (2019) | SNS exhaustion | Adopted MBI for social media exhaustion | Survey study Structural equation modeling | 1285 SNS users (university staff and students ) | Social overload increases SNS exhaustion among all users (emotionally stable, emotionally unstable, lonely and not lonely) Social support on SNS decreases SNS exhaustion only among the emotionally stable and lonely users | ß =.25, ß =.30, ß =.27, ß =.38 ß =−0.14, ß =−0.14 |
| #26 Luqman et al. (2017) | SNS exhaustion | Adopted MBI for social media exhaustion | Survey study Structural equation modeling | 360 SNS users (students) | Excessive social use of SNS increases SNS exhaustion Excessive hedonic use of SNS increases SNS exhaustion Excessive cognitive use (=information overload) of SNS increases SNS exhaustion | ß =.17 ß =.14 ß =.27 |
| #27 Maier et al. (2015) | SNS exhaustion | Adopted MBI for social media exhaustion | Survey study Structural equation modeling | 571 SNS users | Extent of SNS usage; the number of SNS friends, and the subjective social support norm all increase SNS exhaustion through social overload | ß = .24 ß = .12 ß = .46 ß = .62 |
| #28 Myrick (2015) | Energy (mood) depletion (mood) | 1 adjective item questions about emotional state prior and post watching videos | Survey study t-test | 6795 Internet users who watch cat videos | Watching cat videos increases energy and decreases depletion | Cohen's d =.37 Cohen's d =.96 |
| #29 Piszczek (2017) | Work-related emotional exhaustion | MBI | Longitudinal study: 2 wave survey study (1 month apart), emotional exhaustion was assessed at T2 Regression analysis | 163 alumni of a human resource management master's degree program | After hours work-related cell-phone use expectations (from employer) increase actual ICT use and emotional exhaustion directly. After hours work-related cell-phone use influence exhaustion indirectly through perceived boundary control (PBC lowers exhaustion). | B = .57 B= .34 B= .87 B= −.55 |

| Study Number | Subjective Energy Term | Instrument | Methodology, Method | Sample | Main Findings | Effect size |
|---|---|---|---|---|---|---|
| #29 Piszczek (2017) | Work-related emotional exhaustion | MBI | Longitudinal study: 2 wave survey study (1 month apart), emotional exhaustion was assessed at T2\n\nRegression analysis | 163 alumni of a human resource management master's degree program | After hours work-related cell-The relationship is moderated by work-family segregation preferences. At high (low) levels of work–family segmentation preferences there is an indirect, positive (negative) relationship between work–family technology use and emotional exhaustion through boundary control.\nNote that there are more employees in the sample who prefer to segregate their work-family life than to integrate it (M=3.5, SD=.98) and that expectations lower PBC. | Coeff. = .13 (-.15)\n\n\n\n\n\nB=-.32 |
| #30 Quinones and Griffiths (2017) | Energy (recovery, mood) | Self-developed 3 item recovery survey; momentary experience of energy | Longitudinal study: 3 times-a-day survey for 4 consecutive days\n\nMultilevel mixed model analysis | 84 employees | Compulsive Internet use at work (on the day) decrease energy (recovery) before going to bed and for the more compulsive users, increases it for less compulsive users\n\nThe effect persist in the morning after.\n\nCompulsive Internet before bed decrease energy (recovery) before going to bed for the more compulsive users, does not effect for less compulsive users | Interaction:\nß = −.39\nsimple slopes:\nB=-1.03, SD=.36\nB=1.35,SD=.47\n\nInteraction:\nß = −.52\n\n\nInteraction:\nß = −.35\nsimple slopes:\nB=-1.18, SD=.36\nB=0.90,SD=.54 |
| #31 Ragsdale and Hoover (2016) | Work-related emotional exhaustion | MBI | Longitudinal study: 2-wave survey study: predictors at Time 1, criteria at Time 2 (after one week)\n\nHierarchical regression | 213 full time employees, cell-phone users | Work-related cell-phone use (WRCPU) increases emotional exhaustion only for those low on "cell-phone attachment" (CPA; those who answer immediately and cannot imagine their lives without a phone). | ß = −.37\nß = −1.38\n(WRCPU x CPA) |

| Study Number | Subjective Energy Term | Instrument | Methodology, Method | Sample | Main Findings | Effect size |
|---|---|---|---|---|---|---|
| #31 Ragsdale and Hoover (2016) | Work related vigor (as part of work engagement) | UWES | Longitudinal study: 2-wave survey study: predictors at Time 1, criteria at Time 2 (after one week)<br><br>Hierarchical regression | 213 full time employees, cell-phone users | WRCPU increases work-family conflict for both groups (low and high on CPA). Conflict is related positively to exhaustion<br><br>WRCPU increases vigor (as part of work engagement) only for those high on CPA | ß=.42, ß =.42<br><br>ß =.21 (WRCPU x CPA)<br>ß = 1.57 |
| #32 Rashid and Asghar (2016) | Student vigor (as part of engagement) | UWES | Survey study<br><br>Path analysis, Regression analysis | 761 female undergraduate students | Technology use (in general) increases student engagement directly, and through self-directed learning<br><br>Specifically, Internet use, email use and social media use increase student engagement, video gaming decrease it<br>There was no effect found for cell-phone use (texting and calling and app use), media sharing and watching TV | ß = 0.31, ß = 0.32, ß = 0.45<br><br>ß = .14, ß = 0.19 and ß = 0.14, ß = −.13 |
| #33 Rhee and Kim (2016) | Energy (mood)<br><br>Fatigue (mood)<br><br>Work-related emotional exhaustion<br><br>Work-related vigor | AD ACL<br><br><br><br>MBI<br><br><br>UWES | Survey study<br><br>Structural equation modeling | 425 employees | Type of break – smart-phone breaks(sb) vs. conventional break (cb) –moderate the relationship between psychological detachment and fatigue<br>Type of breaks do not moderate the relationship between psychological detachment and energy<br>Energy is related to after-work vigor (no moderation)<br>Fatigue is related to emotional exhaustion (no moderation)<br>The average level of emotional exhaustion was significantly higher in the smart phone group than in the conventional group | ß = −.19 for sb; ß = .04 for cb<br><br>ß =.53 for sb ß = .42 for cb<br><br><br>ß =.57 for sb ß = .66 for cb<br><br>ß =.46 for sb ß = .30 for cb<br><br>NR |

| Study Number | Subjective Energy Term | Instrument | Methodology, Method | Sample | Main Findings | Effect size |
|---|---|---|---|---|---|---|
| #4 Salanova, Grau, Cifre, and Llorens (2000) | Work-related emotional exhaustion | MBI | Survey study<br><br>Hierarchical regression analysis | 140 workers from five different companies form the tile sector and public administration | Computer training increases exhaustion for those low on self-efficacy; it decreases it for those high on self-efficacy | B=-91; training x self-efficacy B=-.93 |
| #35 Sardeshmukh et al. (2012) | Work-related emotional exhaustion<br><br>Vigor (as part of job engagement) | MBI<br><br>Britt's (1999) job engagement scale | Survey study<br><br>Structural equation modeling | 417 employees from a large supply chain management company | Extent of telework decreases exhaustion trough reducing time pressure and role conflicts, and through increasing autonomy<br><br>Extent of telework increases exhaustion trough increasing role ambiguity and through decreasing feedback and social support<br><br>Extent of telework increases engagement trough increasing autonomy<br><br>Extent of telework decreases engagement trough decreasing feedback and social support | ß = −.10; ß = .09; ß = −.16; ß = .14 ß = .12; ß = −.34<br><br>ß =.13; ß = .16 ß = −.22; ß = −.11 ß = −.10; ß =−.28<br><br>ß = .12; ß = .30<br><br>ß = −.22; ß = .23 ß = −.10; ß = .19 |
| #36 Satici and Uysal (2015) | Subjective vitality | SVS | Survey study<br><br>Regression analysis | 311 university students from 2 mid-sized universities | Problematic Facebook use is negatively related to vitality | ß = −.15 to ß = −.24 (depending on other covariates in the models) Single bivariate correlation r=-.32 |
| #37 Shin and Shin (2016) | Mobile messenger fatigue | Adopted the scale from A. R. Lee et al. (2016) | Survey study<br><br>Structural equation modeling | 334 mobile messenger users | Mobile messenger overload (commercial and non-commercial) is related to messenger fatigue. Personality (relational self) moderates the non-commercial relationship | ß = .12; ß = .63<br><br>ß = .67; ß = .69 |
|  |  |  |  |  |  |  |

| Study Number | Subjective Energy Term | Instrument | Methodology, Method | Sample | Main Findings | Effect size |
|---|---|---|---|---|---|---|
| #38 van Zoonen and Rice (2017) | Work-related emotional exhaustion  Work-related vigor | MBI  UWES | Survey study  Structural equation modeling | 364 employees; 102 from a Telecom provider; 112 from a consultancy firm and 150 from a consumer electronics company | Using SNS for work purposes (Twitter, LinkedIn and Facebook) increases exhaustion through increasing work pressure  Using SNS for work purposes (Twitter, LinkedIn and Facebook) decreases exhaustion through increasing autonomy  Responsiveness moderates the relationship between SNS use and autonomy, i.e., the above benefits of autonomy are only valid for those low in responsiveness.  Using SNS for work purposes (Twitter, LinkedIn and Facebook) increases vigor through increasing autonomy (the moderation impacts this relationship, too). | ß = .48, ß = .43  ß = .15; ß = −.25  ß = .08  ß = .15, ß = .37 |
| #39 Windeler, Chudoba, and Sundrup (2017) #40 Windeler, Chudoba, and Sundrup (2017) | Work-related emotional exhaustion | MBI | Study 1: Longitudinal, 2-wave survey study: 1 week prior to-telework including baseline exhaustion and 4 months later (post telework);  Structural equation modeling | 51 employees from a IT business unit of a financial service firm randomly selected to part-time telework | Part time teleworking (PTT) weakness the positive relationship between interpersonal interaction and work exhaustion but strengthens the positive relationship between external interaction and work exhaustion. PTT does not moderate the relationship between interdependence and exhaustion | ß = −.11 pre; ß = .37 post teleworking ß = .28 pre; ß = .03 post teleworking ß = .19 both pre and post teleworking |

| Study Number | Subjective Energy Term | Instrument | Methodology, Method | Sample | Main Findings | Effect size |
|---|---|---|---|---|---|---|
| #39 Windeler, Chudoba, and Sundrup (2017) #40 Windeler, Chudoba, and Sundrup (2017) | Work-related emotional exhaustion | MBI | Study 2: Survey study; Structural equation modeling | 258 employees (160 part time teleworkers. Note. employees classified as no part time teleworkers had the option to telework but chose not to) | Job interdependence explains more of the exhaustion variance for no part time employees than for PT teleworkers External interaction explained less of the exhaustion variance for no part time employees than for PT teleworkers Interpersonal interaction quantity explained more of the exhaustion variance for no part time employees than for PT teleworkers Interpersonal interaction quality had no impact on the relationship | ß = .31 for No PTT, ß = .17 for PTT ß = .18 for No PTT, ß = .38 for PTT ß = .19 for No PTT, ß = .05 for PTT ß = −..34 for No PTT, ß = −..25 for PTT (ns.) |
| #40 Xie, Ma, Zhou, and Tang (2018) | Work-related emotional exhaustion | MBI | Study 1: Survey study Regression analysis | 447 college councilors | Study 1. Work related ICTs use after work hours increases exhaustion even after controlling for integration preference. Integration preference is negatively related to exhaustion (even after controlling for ICT use). The interaction is also significant predictor: use increases exhaustion for those low in integration preferences, but not for those high in integration preferences. Note that there are more employees in the sample who prefer to segregate their work-family life than to integrate it (M=2.15, SD=1.08, 1-5 Likert, higher values = higher integration preferences.) | ß = .24 ß = −.26 ß (interaction) = −.20 ß = .42 for low integration ß=.07 for high integration preferences |

| Study Number | Subjective Energy Term | Instrument | Methodology, Method | Sample | Main Findings | Effect size |
|---|---|---|---|---|---|---|
| #40 Xie, Ma, Zhou, and Tang (2018) | Work-related emotional exhaustion | MBI | Study 2: Survey study <br><br> Regression analysis | 437 full time employees | Study 2. Work related ICTs use after work hours increases exhaustion even after controlling for integration preference. The interaction between ICT use and integration preference is also a significant predictor Autonomy (work schedule and location control) mediates the relationship (lowers exhaustion) Note that there are more employees in the sample who prefer to segregate their work-family life than to integrate it (M=2.05, SD=1.04, 1-5 Likert, higher values = higher integration preferences.) | ß = .26 <br> ß = .28 <br> ß = −.16 |
| #41 Zhang et al. (2016) | SNS fatigue | Adopted and self-developed scale for SNS fatigue | Survey study <br><br> Regression analysis | 525 SNS (Qzone) users | System feature overload, information overload and social overload drive SNS fatigue | ß = .24 <br> ß = .20 <br> ß = .37 |
| #42 Zheng and Lee (2016) | SNS exhaustion (strain) | Adopted MBI for social media exhaustion | Survey study <br><br> Regression analysis | 550 mobile SNS users | Excessive SNS use and cognitive preoccupations increase SNS exhaustion through creating technology-family conflict; technology-personal conflict and technology-work conflict (technology-personal conflict is predicted additionally both by technology-family and technology-work conflict) | ß = .60, ß = .08 <br> ß = −.08, ß = .65 <br> ß = .42, ß = .23 <br> ß = .11, ß = .08 <br> ß = .18, ß = .65 <br> ß = .17, ß = .23 <br> ß = .35, ß = .45 |